\documentclass[twocolumn,english,aps,prx,superscriptaddress]{revtex4-1}
\usepackage[latin9]{inputenc}
\usepackage{amsmath}
\usepackage{graphicx}
\usepackage{color}
\usepackage{esint}
\usepackage{bbold}

\makeatletter

\providecommand{\tabularnewline}{\\}

\usepackage{physics}

\makeatother

\usepackage{babel}

\begin{document}

\title{Classification of crystalline insulators without symmetry indicators: \\
atomic and fragile topological phases in  twofold rotation symmetric systems}

\author{Sander H. Kooi}
\affiliation{Institute for Theoretical Physics, Center for Extreme Matter and
Emergent Phenomena,\\ Utrecht University, Princetonplein 5, 3584
CC Utrecht, the Netherlands}

\author{Guido van Miert}
\affiliation{Institute for Theoretical Physics, Center for Extreme Matter and
Emergent Phenomena,\\ Utrecht University, Princetonplein 5, 3584
CC Utrecht, the Netherlands}

\author{Carmine Ortix}
\affiliation{Institute for Theoretical Physics, Center for Extreme Matter and
Emergent Phenomena,\\ Utrecht University, Princetonplein 5, 3584
CC Utrecht, the Netherlands}
\affiliation{Dipartimento di Fisica ``E. R. Caianiello'',
Universit\`a di Salerno I-84084 Fisciano (Salerno), Italy }

\date{\today} 

\begin{abstract}
Topological crystalline phases in electronic structures can be generally classified using the spatial symmetry characters of the valence bands and mapping them onto appropriate symmetry indicators. 
These mappings have been recently applied to identify thousands of topological electronic materials.
There can exist, however, topological crystalline non-trivial phases that go beyond this paradigm: they cannot be identified using spatial symmetry labels and consequently lack any classification. 
In this work, 
we achieve the first of such classifications showcasing
the paradigmatic example of two-dimensional crystals with twofold rotation symmetry. 
We classify the gapped phases in time-reversal invariant systems with strong spin-orbit coupling identifying a set of three $\mathbb{Z}_2$ topological invariants, which correspond to nested quantized partial Berry phases. By further isolating the set of atomic insulators representable in terms of exponentially localized symmetric Wannier functions, we infer the existence of topological crystalline phases of the fragile type that would be diagnosed as topologically trivial using symmetry indicators, and 
construct a number of microscopic models
exhibiting this phase. Our work is expected to have important consequences given the central role fragile topological phases are expected to play in novel two-dimensional materials such as twisted bilayer graphene. 

\end{abstract}

\maketitle

\section{Introduction}
Since the discovery of the quantum Hall effect \cite{kli80}, and its theoretical explanation in terms of the topological properties of the Landau levels \cite{tho82,hal82,koh85}, topological phases of matter have become a rich playground for the theoretical prediction and experimental verification of new quantum phenomena. 
From the birth of topological insulators \cite{kan05b,kan05,ber06,mol07,zha09,fu07,bru11,ras13}, to topological superconductors supporting Majorana zero modes \cite{mou12,lut10,fu08,bee13,ali12}, to topological semimetals\cite{arm18,hua15,lv15,xu15,hau17,lau17,wan11,bur11,zyu12,lau19,oja13,sol15}, new types of topological phases keep arising. 
It is fair to say that the major theoretical effort in the field has been to classify, using appropriate mathematical schemes, all possible topologically distinct gapped phases and subsequently relate them to topological indices. 
In the presence of internal symmetries -- time-reversal, particle-hole and chiral symmetry -- alone, the classification of  free-fermion gapped phases has been obtained in all ten symmetry classes and arbitrary number of dimensions \cite{alt97,sch08,kit09}. The corresponding phases with non-trivial topology feature, by the bulk-boundary correspondence, protected gapless modes that are anomalous \cite{has10,qi11}. The chiral (helical) edge states of quantum (spin) Hall insulators, as well as the single surface Dirac cones of strong three-dimensional topological insulators  violating the fermion doubling theorem, are prime realizations of such anomalies. 

In crystalline systems characterized by an additional set of spatial symmetries, new topologically distinct phases emerge \cite{fu11,hsi12,liu14b,hsi14}. The non-trivial topology of a system is then manifested in the appearance of anomalous gapless surface modes, which are present only on surfaces that are left invariant under the protecting spatial symmetry and violate stronger versions of the fermion doubling theorem \cite{fan17,kha18}. Furthermore, crystalline symmetries  can also yield non-trivial topological phases, dubbed higher-order topological states \cite{sch18,ben17,ben17-2,sch18b,gei18,pet18,ser18,kha18b,eza18,eza18b,eza18,lan17,sit12,son17,imh18,xu17,hsu18,mie18,koo18}, characterized by conventional gapped surfaces but with gapless anomalous one-dimensional modes at the hinges connecting two surfaces related by the protecting spatial symmetry. 

As long as insulating systems are concerned, the existence of anomalous surface or hinge boundary modes is deeply connected to the fact that non-trivial topological phases cannot be adiabatically connected to atomic insulators, whose insulating nature can be understood considering electrons as trapped classical point particles. In other words, a topological non-trivial insulator only arises when there is an obstruction in describing the system using an atomic picture. Therefore, the ground state of a topological non-trivial insulator cannot be represented using exponentially localized Wannier functions respecting the internal and/or the set of spatial symmetries of the system \cite{sol11}. This obstruction to a ``Wannier-representability", the classification in terms of topological invariants and the existence of gapless anomalous boundary modes can be formulated in a unique consistent framework for systems equipped only with internal symmetries \cite{yu11,sol11}. When adding spatial symmetries, however, 
different complications arise.

First, distinct atomic insulators, which are by definition topologically trivial, 
generally possess different crystalline topological invariants. This, in turn, requires a careful inspection of such topological indices to identify the criteria dictating the appearance of topologically non-trivial crystalline phases. 
Second, there can exist ``non-Wannierazible" topological phases in crystals which do not possess boundaries that are left invariant under the protecting spatial symmetry.
 As a result, the surfaces of these systems are fully gapped even if the bulk is  topological. 
Notwithstanding these complications, substantial progresses has been made with
the theory of topological quantum chemistry \cite{bra17} and that of symmetry-based indicators \cite{po17,kru17b,son18}, which allows one to discriminate all different atomic insulators from genuine topological non-trivial phases using the spatial symmetry character of the valence bands and their connectivity throughout the Brillouin zone. Combining these theories with density-functional-theory calculations has very recently led to catalogues containing a huge number of topological materials \cite{zha19,ver19,tan19}.

Nonetheless, there exist topological phases that are not detectable using the symmetry labels of the valence bands. An extreme case is a system with only translation symmetry: it can be in a topological ``tenfold-way" phase due to its internal symmetries, but it is signaled as being topologically trivial using spatial symmetry indicators. More importantly, there can exist topological crystalline phases in low-symmetric crystals that are neither characterizable by the symmetry content of the valence bands nor by the tenfold-way \cite{son19}. To date, these phases lack any classification and consequently any  material realization. 

In this work, we achieve the first of such classifications.
Specifically, we consider the paradigmatic example of two-dimensional crystals with twofold rotation symmetry, {\it i.e.} in the wallpaper group $p2$, where the gapped phases of time-reversal symmetric (non-magnetic) systems with sizable spin-orbit coupling cannot be classified with the symmetry data of the valence bands. Instead, we construct Berry phase related ${\mathbb Z}_2$ invariants to first isolate and remove 
topologically non-trivial quantum spin-Hall phases from the set of distinct gapped phases. Thereafter, we enumerate all distinct atomic insulating phases and classify them using a trio of ${\mathbb Z}_2$ topological invariants. Using our Berry phase based classification, we are able to determine: {\it i)} in systems with two occupied valence bands, the existence of topological non-trivial crystalline phases similar in nature to the fragile phases detected by symmetry eigenvalues in 
other wallpaper groups~\cite{po18,bra19}. 
{\it ii)} with four occupied valence bands, the emergence of an additional fragile topological crystalline phase, whose possible existence has been overlooked so far. To underline the importance of these findings, we point out that topological crystalline phases of the fragile type have 
been predicted to occur in magic-angle twisted bilayer graphene \cite{cao2018,cao18b,ahn19,po19}.

This paper is organized as follows. In Sec.~II we first present the example of a time-reversal symmetric one-dimensional atomic chain where the symmetry character of the bands is not able to classify the distinct gapped phases, and show that such a classification becomes instead possible introducing a ``partial" Berry phase $\mathbb{Z}_2$ invariant. 
We then show in Sec.~III that these $\mathbb{Z}_2$ invariants can be also defined on high-symmetry lines in the Brillouin zone of a two-dimensional crystal in the $p2$ wallpaper group, and can be used to first remove topological phases protected by time-reversal symmetry, and then classify atomic and fragile topological phases when two valence bands are occupied. In Sec.~IV we introduce a new $\mathbb{Z}_2$ invariant corresponding to a ``nested" quantized partial Berry phase, thanks to which we are able to diagnose the atomic insulating phases realized with four occupied valence bands and establish the existence of our novel $N_F=4$ fragile topological insulator. The trio of $\mathbb{Z}_2$ invariants is then used to classify all atomic insulating phases for a generic number of occupied Kramers pairs of bands in Sec.~IV.  Finally, we present our conclusions and comment on extensions of our work in Sec.~V.

\section{Motivation and warmup in 1D: mirror-symmetric chains} 
We start out by considering 
an atomic chain of spin one-half electrons with time-reversal symmetry and an additional mirror symmetry with respect to a one-dimensional (1D) mirror point. Moreover, we will assume inversion symmetry to be explicitly broken. 
The space group ${\mathcal G}$ for this atomic chain is generated by 
$${\mathcal G}= \langle \left\{E | {\bf t} \right\}, \left\{{\mathcal M} | 0 \right\} \rangle, $$
where $E$ is the identity, ${\bf t}$ the lattice translation vector, and ${\mathcal M}$ the mirror symmetry with respect to the 1D mirror point. 
In the unit cell of this 1D crystal, there are two distinct maximal Wyckoff positions whose site symmetry group, or stabilizer group, is isomorphic to the point group ${\mathcal C}_s$.  
The first, labelled  $1a$, has coordinate $x=0$ and corresponds to the origin of the unit cell. Its stabilizer group is simply generated by $\left\{{\mathcal M} | 0 \right\}$. Similarly, the second maximal Wyckoff position, labelled $1b$, corresponds to the edge of the unit cell with coordinate $x=1/2$ in units of the lattice constant, and its stabilizer group is generated by $\left\{{\mathcal M} | 1 \right\}$, which is also isomorphic to ${\mathcal C}_s$.
For all other positions in the unit cell, the stabilizer group only contains the identity. Therefore these Wyckoff positions have multiplicity two and coordinates $\left(x,-x\right)$. Let us now enumerate the elementary band representations~\cite{zak82} for exponentially localized Wannier functions (WFs) sitting at the maximal Wyckoff positions $1a$ and $1b$. 
They can be induced by considering that in reciprocal space there are two mirror-symmetric momenta in the Brillouin zone (BZ), {\it i.e.} $\Gamma=0$ and $X=\pi$. Moreover, since the stabilizer group of $1a$ does not contain any translation, the mirror eigenvalues $\pm i$ at $\Gamma$ and $X$ must be identical. On the contrary, the stabilizer group of $1b$ contains a lattice translation of half a unit cell and therefore the mirror eigenvalues at $\Gamma$ and $X$ are opposite. The elementary band representations can then be summarized as in Table~\ref{tab:tab1d}. 
Note that the ``composite" band representation for two symmetric WFs~\cite{mie18} at the same position with opposite mirror eigenvalues $\pm i$ have a representation content in momentum space that is independent on whether they are centered at $1a$ or $1b$. This yields the
equivalence $\rho_i^{1a} \oplus \rho_{-i}^{1a} \uparrow {\mathcal G} \simeq \rho_i^{1b} \oplus \rho_{-i}^{1b} \uparrow {\mathcal G}$, which simply states that the corresponding pairs of exponentially localized WFs can be moved 
anywhere along the line between the $1a$ and the $1b$ sites in opposite directions. 

\begin{table}
\begin{tabular}{|c|c|c|c|}
\hline 
Wyckoff position & Representation & $\Gamma$ & $X$ \tabularnewline
\hline 
$1a$ & $\rho_{i}^{1a} \uparrow {\mathcal G}$ & $i$ & $i$  \tabularnewline 
& $\rho_{-i}^{1a} \uparrow {\mathcal G}$ & $-i$ & $-i$  \tabularnewline 
\hline 
$1b$ & $\rho_{i}^{1b} \uparrow {\mathcal G}$ & $i$ & $-i$  \tabularnewline 
& $\rho_{-i}^{1b} \uparrow {\mathcal G}$ & $-i$ & $i$  \tabularnewline 
\hline 
\end{tabular}
\caption{Elementary band representation for the one-dimensional space group of a mirror symmetric chain. The first column indicates the maximal Wyckoff positions. The second column the corresponding induced band representation, and the last two columns the mirror eigenvalues at the center and edge of the 1D BZ.}
\label{tab:tab1d} 
\end{table} 

The aforementioned composite band representation becomes a physical  elementary band representation (PEBR)~\cite{bra17}
when time-reversal symmetry $\Theta$ is taken into account. This is because $\Theta$ requires the 
complex irreducible one-dimensional representations at $\Gamma$ and $X$ to double. 
The corresponding pairs of energy bands, however, do not derive from Wannier states with charge centers at arbitrary positions along the chain. Kramers theorem indeed guarantees that exponentially localized WFs come in Kramers degenerate pairs, in which each pair has the same center. 
Moreover, while an even number of Wannier Kramers pairs centered at the maximal Wyckoff positions $1a$ or $1b$ can be freely moved away without breaking either the mirror or time-reversal symmetry, with an odd number of Wannier Kramers pairs sitting at $1a$ or $1b$ the center of at least one pair of Wannier states is unmovable~\cite{son17}.
Put differently, the parity of Wannier Kramers pairs centered at the maximal Wyckoff positions $1a$ and $1b$ represent stable topological $\mathbb{Z}_2$ indices characterizing a one-dimensional time-reversal and mirror-symmetric insulator. More importantly, these stable topological indices cannot be read off from the symmetry character of the bands since 
only one PEBR exists. 
The discrepancy between the existence of real space stable topological indices and the absence of distinct PEBRs can be overcome using the recent finding that Kramers pairs of bands in a mirror symmetric~\cite{lau16}, or equivalently  ${\mathcal C}_2$ twofold rotation symmetric~\cite{mie17},  atomic chain possess a  ${\mathbb Z}_2$ topological index defined in terms of the ``partial" polarization introduced by Fu and Kane~\cite{fu06}, which is quantized by the presence of these point group symmetries. 
In its $U(N_F)$ gauge invariant form it can be written as 
\begin{equation}
\nu^{{\mathcal M}}:= \dfrac{1}{\pi} \left[ \int_0^{\pi} d k \, \textrm{Tr} \,  {\mathcal A}(k) + i \log \dfrac{ \textrm{Pf} \left[w(\pi)\right]}{ \textrm{Pf} \left[w(0)\right]}  \right] \, \textrm{mod}~2. 
\label{eq:partialpol}
\end{equation}
In the equation above, we have introduced the non-Abelian Berry connection ${\mathcal A}_{m,n}(k)=\bra{u_m(k)} i \partial_k \ket{u_n(k)}$, and the sewing matrix $w_{m,n}(k)=\bra{u_m(-k)} \Theta \ket{u_n(k)}$ that is antisymmetric at the $\Gamma$ and $X$ points and hence characterized by its Pfaffian Pf$(w)$. 
The $\mathbb{Z}_2$ invariant defined above can be related to the charge centers of the Wannier Kramers pairs by introducing the unitary  Wilson loop operator~\cite{yu11,ale14}
\begin{equation}
{\mathcal W}_{k+2\pi \leftarrow k}=\overline{\mathrm{exp}} \left[ i \int_k^{k + 2 \pi} {\mathcal A}(k^{\prime}) d k^{\prime} \right], 
\end{equation}
where $\overline{\mathrm{exp}}$ denotes path ordering of the exponential while $k$ is the Wilson loop base point. The eigenvalues of the Wilson loop operator,  $\mathrm{exp}(2 \pi i \, \nu_j)$, $j$ labelling the occupied bands, are independent of the base point $k$ and uniquely determine the Wannier centers $\nu_j$. 
The presence of mirror symmetry translates into a chiral symmetry for the Wilson loop eigenvalues~\cite{bra19}, thus implying that the Wannier centers are restricted to the values $\nu_j=0,1/2$ or to ``unpinned" pairs $(\bar{\nu}, -\bar{\nu})$. Moreover, time-reversal symmetry guarantees that each Wilson loop eigenvalue has to be doubly degenerate. The concomitant presence of mirror and time-reversal symmetry therefore yields $\sum_j \nu_j ~\textrm{mod}~1 \equiv 0$, and consequently $\sum_j \nu_j ~\textrm{mod}~2 \equiv \nu^{{\mathcal M}}$ can only assume the values $0$ and $1$. 
Knowing the relation between the $\mathbb{Z}_2$ topological invariant and the Wannier centers, we can straightforwardly classify the insulating states realized in a one-dimensional mirror-symmetric atomic chain. In fact,  with a total number of occupied bands $N_F=4n + 2$, $n$ being integer, an insulating atomic chain for which $\nu^{\mathcal M}=0$ ($\nu^{\mathcal M}=1$) will be characterized by the presence of an odd number of Wannier Kramers pairs at $1a$ ($1b$). If instead $N_F=4n$ the system can be described in terms of exponentially localized Wannier functions with an even or odd number of Kramers degenerate pairs centered at $1a$ \emph{and} $1b$ depending on whether $\nu^{\mathcal M}=0$ or $\nu^{\mathcal M}=1$, respectively. 

\begin{figure}
\includegraphics[width=\columnwidth]{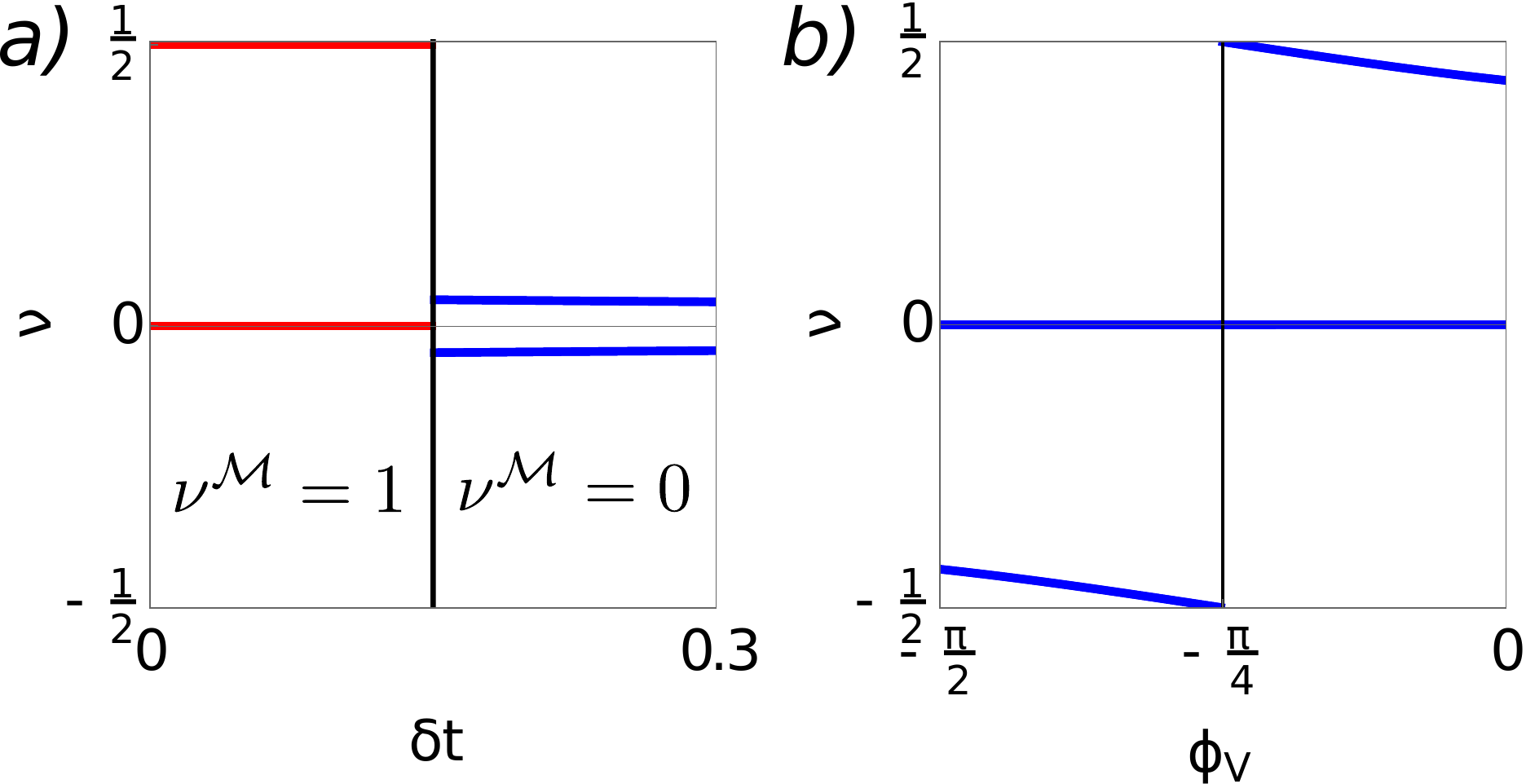}
\caption{(color online) Evolution of the Wilson loop eigenvalues for a mirror and time-reversal symmetric Aubry-Andr\'e-Harper model [cf. Appendix \ref{sec:harper} and  Ref.~\onlinecite{lau16}] at $N_F=4$ by sweeping the dimerization hopping amplitude $\delta t$ while preserving mirror symmetry (a) and changing the phase $\phi_V$ away from the mirror-symmetric point $\phi_V=-\pi/4$ (b).}
\label{fig:fig1}
\end{figure}

To verify the relation between the $\mathbb{Z}_2$ topological invariant $\nu^{\mathcal M}$ and the Wannier centers distribution, we have computed the Wilson loop spectrum for a time-reversal and mirror symmetric one-dimensional spinful Aubry-Andr\'e-Harper model [cf. Appendix A and Ref.~\onlinecite{lau16}], in which the half-filled $N_F=4$ insulating state undergoes a band gap closing-reopening, accompanied by a change of the $\mathbb{Z}_2$ topological invariant, by sweeping the strength of the nearest-neighbor hopping amplitude $\delta t$. As explicitly shown in Fig.~\ref{fig:fig1}(a), the insulating state can be described in terms of two Wannier Kramers pairs centered at $1a$ and $1b$  in the $\nu^{\mathcal M}=1$ region.  On the contrary, a $\nu^{\mathcal M}=0$ value of the topological invariant implies the existence of two Wannier pairs centered at two mirror related, non-maximal Wyckoff positions in the unit cell.  Moreover, by breaking the mirror symmetry of the model [see Fig.~\ref{fig:fig1}(b)] the position of the exponentially localized Wannier function can be freely moved at arbitrary positions in the unit cell in agreement with the fact that the space group in this case only contains the identity. Finally, 
we emphasize that the change of the $\mathbb{Z}_2$ invariant is associated with a band gap closing-reopening occurring at unpinned points in the BZ~\cite{lau16}, which is a restatement of the fact that the topological index characterizing a mirror and time-reversal symmetric insulating chain cannot be inferred from the symmetry character of the occupied bands. 

\begin{table}
\begin{tabular}{|c|c|c|c|c|c|}
\hline 
Wyckoff position & Representation & $\Gamma$ & $X$ & $Y$ & $M$  \tabularnewline
\hline 
$1a$ & $\rho_{i}^{1a} \uparrow {\mathcal G}$ & $i$ & $i$ & $i$ & $i$  \tabularnewline 
& $\rho_{-i}^{1a} \uparrow {\mathcal G}$ & $-i$ & $-i$  & $-i$ & $-i$ \tabularnewline 
\hline 
$1b$ & $\rho_{i}^{1b} \uparrow {\mathcal G}$ & $i$ & $-i$ & $i$ & $-i$  \tabularnewline 
& $\rho_{-i}^{1b} \uparrow {\mathcal G}$ & $-i$ & $i$ & $-i$ & $i$  \tabularnewline 
\hline 
$1c$ & $\rho_{i}^{1c} \uparrow {\mathcal G}$ & $i$ & $i$ & $-i$ & $-i$  \tabularnewline 
& $\rho_{-i}^{1c} \uparrow {\mathcal G}$ & $-i$ & $-i$ & $i$ & $i$  \tabularnewline 
\hline 
$1d$ & $\rho_{i}^{1d} \uparrow {\mathcal G}$ & $i$ & $-i$ & $-i$ & $i$  \tabularnewline 
& $\rho_{-i}^{1d} \uparrow {\mathcal G}$ & $-i$ & $i$ & $i$ & $-i$  \tabularnewline 
\hline 
\end{tabular}
\caption{Elementary band representation for the $p2$ wallpaper group ${\mathcal G}= \langle \left\{E | {\bf t} \right\}, \left\{{\mathcal C}_2 | 0 \right\} \rangle$. The first column indicates the maximal Wyckoff positions; the second column the corresponding induced band representation, and the last two columns the $\mathcal{C}_2$ eigenvalues at the $\Gamma=\left\{0,0\right\}$, $X=\left\{\pi,0\right\}$, $Y=\left\{0,\pi\right\}$ and $M=\left\{\pi,\pi\right\}$ points in the BZ. In time-reversal symmetric systems, the PEBRs obey the equivalence $\rho^{1a} \uparrow {\mathcal G} \simeq \rho^{1b} \uparrow {\mathcal G} \simeq \rho^{1c} \uparrow {\mathcal G} \simeq \rho^{1d} \uparrow {\mathcal G}$.}
\label{tab:tab2d} 
\end{table} 

\section{Wallpaper group $p2$: insulators with two occupied bands}
Having established the $\mathbb{Z}_2$ classification of mirror and time-reversal symmetric insulating chains  in the absence of symmetry indicators,  
we next consider the main focus of this work: two-dimensional (2D) crystals
possessing a ${\mathcal C}_2$ twofold rotation symmetry. The smallest two-dimensional wallpaper group containing ${\mathcal C}_2$ is $p2$. It has four maximal Wyckoff positions labelled as $1a=\left\{0,0\right\}$, $1b=\left\{1/2,0\right\}$, $1c=\left\{0,1/2\right\}$ and $1d=\left\{1/2,1/2\right\}$. Their stabilizer group is isomorphic to ${\mathcal C}_2$, which implies that in systems with time-reversal symmetry the induced band representations have the same symmetry character [cf. Table~\ref{tab:tab2d}]. 

\begin{figure}
\begin{centering}
\includegraphics[width=.9\columnwidth]{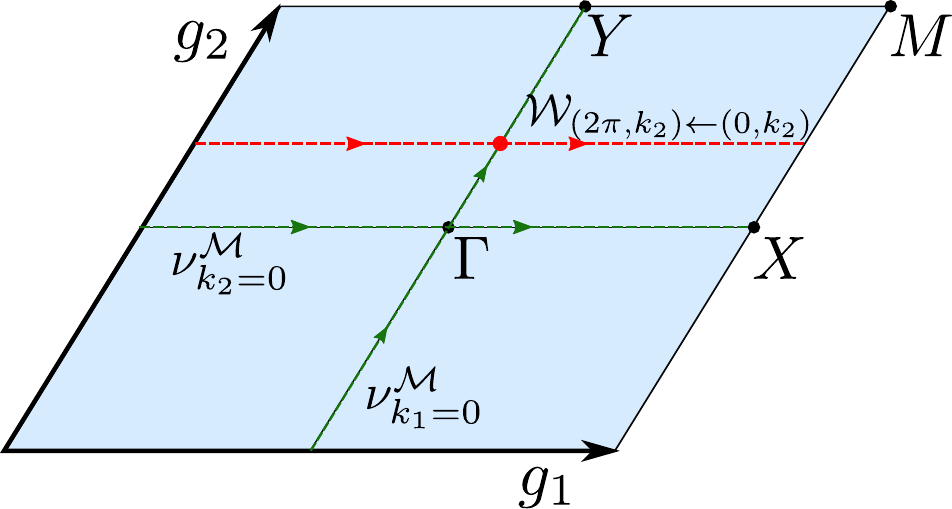}
\par\end{centering}

\caption{(color online) Schematic drawing of a $\mathcal{C}_{2}$ symmetric Brillouin zone spanned by reciprocal lattice vectors $g_{1}$ and $g_{2}$ with high-symmetry points $\Gamma$, $X$, $Y$ and $M$. The contours along which the partial Berry phases $\gamma_{1}^{I}$ and $\gamma_{2}^{I}$ are calculated are drawn in green, a typical Wilson loop operator contour, discussed in the main text, is drawn in red. \label{fig:bz}}

\end{figure}

However, the parity of the Wannier Kramers pairs centered at $1a$,$1b$,$1c$,$1d$ still represent real space stable topological indices that discriminate between non-equivalent atomic insulating states. To classify these different atomic insulators, we 
first use the fact that
in the BZ of a twofold rotation symmetric crystal, 
the ${\mathcal C}_2$ symmetry constraint ${\mathcal C}_2^{-1} {\mathcal H}({\bf k}) {\mathcal C}_2= {\mathcal H}(-{\bf k})$ 
is equivalent to 
a one-dimensional mirror symmetry constraint along the time-reversal invariant non-contractible loop lines $k_{1,2} \equiv 0$, and $k_{1,2}={\bf G}_{1,2}/2$. Therefore, we can in principle define a quartet of $\mathbb{Z}_2$ invariants $\left\{\nu^{\mathcal M}_{k_{1}=0}; \nu^{\mathcal M}_{k_{1}={\bf G}_1 / 2}; \nu^{\mathcal M}_{k_{2}=0};  \nu^{\mathcal M}_{k_{2}={\bf G}_2 / 2}\right\}$ [c.f. Fig. \ref{fig:bz}]. These topological indices  are not all independent, however, since the differences $ \nu^{\mathcal M}_{k_{1,2}={\bf G}_{1,2} / 2} - \nu^{\mathcal M}_{k_{1,2}=0}$ can be related~\cite{kru17}
 to the Fu-Kane-Mele (FKM) $\mathbb{Z}_2$ topological invariant~\cite{kan05,fu06} characterizing a time-reversal invariant 2D topological insulator. This follows from the fact that $ \nu^{\mathcal M}_{k_{1,2}={\bf G}_{1,2} / 2} - \nu^{\mathcal M}_{k_{1,2}=0}$ keeps track of the evolution of the Wannier centers during a time-reversal pumping process~\cite{yu11}. Therefore, the condition  $ \nu^{\mathcal M}_{k_{1,2}={\bf G}_{1,2} / 2} - \nu^{\mathcal M}_{k_{1,2}=0} = 1~\textrm{mod}~2 $ immediately implies a quantum spin Hall (QSH) insulating state. 
When dealing with insulating crystalline systems without anomalous edge states (trivial FKM invariant), 
we are  thus left with a $\mathbb{Z}_2 \times \mathbb{Z}_2$ classification~\cite{mie18r}, which, as we will show below, is only able to diagnose the atomic insulating states 
when one Kramers pair of bands is occupied. 

The assertion above can be immediately proved by using the fact that for an atomic insulator with two occupied bands, the exponentially localized Wannier Kramars' pair must be centered at one of the maximal Wyckoff positions. Hence, the corresponding center of charge already provides a $\mathbb{Z}_2 \times \mathbb{Z}_2$ classification. Furthermore, the center of charge can be straightforwardly connected to the doublet of one-dimensional invariants $\nu^{\mathcal M}_{k_{1,2}=0}$ as follows. 
 Let us consider the Wilson loop operator in the $e_1$ direction $\mathcal{W}_{(k_1+2\pi, k_2) \leftarrow (k_1,k_2)}$ where $(k_1,k_2)$ is the base point. Its eigenvalues $\mathrm{exp}\left[2 \pi i~\nu_j(k_2)\right]$ ($j=1,2$)  depend on the $k_2$ coordinate of the Wilson loop base point and the corresponding phases $\nu_j(k_2)$ are the centers of the one-dimensional hybrid Wannier functions [c.f. Fig. \ref{fig:bz}]. Due to  time-reversal symmetry the Wannier bands realize a Kramers related pair [c.f. Appendix~\ref{sec:nestedPartial}], and therefore can be split into two time-reversed channels $s=I,II$ satisfying $\nu^{I}(k_2)\equiv \nu^{II} (-k_2)$. The additional ${\mathcal C}_2$ rotation symmetry mandates the Wilson loop spectrum to be chiral symmetric, {\it i.e.} $\nu^{I}(k_2) \equiv -\nu^{II}(k_2)$. As a result, the center of charge of the Wannier Kramers pair in the $e_1$ direction is 
$$\dfrac{1}{2 \pi} \oint \nu^{I} (k_2) d k_2~\textrm{mod}~1  \equiv \nu^{I}(k_2=0)~\textrm{mod}~1 \equiv \dfrac{\nu^{\mathcal M}_{k_2=0}}{2}.$$
Repeating the same argument using the Wilson loop operator in the $e_2$ direction, we therefore reach the classification of atomic insulators with one occupied Kramers pair of bands summarized in Table~\ref{tab:tabNf2}. 

Strictly speaking, this classification does not enumerate all possible insulating phases with a trivial FKM invariant. Contrary to 1D systems where all insulating phases can be adiabatically continued to an atomic insulating phase~\cite{po17}, in 2D systems there can exist topologically non-trivial states that present an obstruction to a representation in terms of symmetric and exponentially localized WFs~\cite{po18}. 
These topological phases have been dubbed ``fragile" topological phases since although not admitting a Wannier representation by themselves, such a representation becomes possible when additional trivial bands are added to the system. 
In recent works, the existence and diagnosis of fragile topological phases~\cite{can18,bra19,po19} have been linked to the topological nature of disconnected PEBR's~\cite{bra17}. However,  the defining characteristic of a fragile topological phase -- the absence of a Wannier gap in the Wilson loop spectrum that consequently must display a non-trivial winding -- can exist also in our low-symmetric crystal with a single unsplittable PEBR. 

\begin{table}[t!]
\begin{tabular}{|c|c|c|}
\hline 
Wyckoff position & $\nu^{\mathcal M}_{k_1=0}$ & $\nu^{\mathcal M}_{k_2=0}$  \tabularnewline
\hline 
$1a$ & 0 & 0 \tabularnewline 
\hline
$1b$ & 0 & 1 \tabularnewline 
\hline 
$1c$ & 1 & 0 \tabularnewline 
\hline
$1d$ & 1 & 1 \tabularnewline
\hline
\end{tabular}
\caption{The ${\mathbb Z}_2 \times {\mathbb Z}_2$ classification of atomic insulators in the $p2$ wallpaper group with one occupied Kramers pair, {\it i.e.} $N_F=2$. The first column indicates the maximal Wyckoff position, while the second and third column are the $U(2)$ gauge invariant line invariants.}
\label{tab:tabNf2} 
\end{table}

In fact, due to the concomitant presence of the commuting two-fold rotation symmetry and time-reversal symmetry, a crystal in the $p2$ space group is also invariant under the combined antiunitary symmetry operation ${\mathcal C}_2 \Theta$ with $({\mathcal C}_2 \Theta)^2=1$. 
Assuming a periodic and smooth real gauge can be found~\footnote{A real gauge can be formulated as $\mathcal{C}_{2}\Theta\ket{\psi}=\ket{\psi}$.}, 
this also implies that the Wilson loop operator in the $e_{1,2}$ direction belongs to the orthogonal group $SO(2)$, with the homotopy group $\pi_1\left[SO(2)\right]=\mathbb{Z}$ guaranteeing the existence of an integer winding number invariant~\cite{bzd17}. A ${\mathcal C}_2 \Theta$-protected fragile topological phase of this kind has been first discussed in Ref.~\onlinecite{ahn18} and dubbed Stiefel-Whitney (SW) insulator since the parity of the winding number corresponds to the second SW class invariant. 
Note that for a SW insulator to exist,  the total Berry phases along the $k_{1,2}\equiv 0$ lines -- which correspond to the first SW class invariant in a smooth and periodic real gauge -- must vanish. This constraint is immediately verified in a ${\mathcal C}_2$ crystal with time-reversal symmetry. On the other hand,  time-reversal symmetry also guarantees the winding number of the Wilson loop operator to assume $2 \mathbb{Z}$ values, which, in the language of Ref.~\onlinecite{ahn18} would imply the $\mathbb{Z}_2$ second SW class invariant to be trivial.  

However, in a $N_F=2$ insulator with time-reversal symmetry a Wilson loop spectrum winding an even number of times cannot be unwinded. Consider the Wilson loop operator $\mathcal{W}_{(k_1, k_2+2 \pi) \leftarrow (k_1,k_2)}$ and assume, for instance, that the line invariant $\nu^{\mathcal M}_{k_1=0}=0$. The Wilson loop spectrum has to display two symmetry enforced degeneracies at $k_1=0,\pi$ with the corresponding hybrid Wannier centers at $\nu=0$. The absence of a Wannier gap also implies the existence of two degeneracies at time-reversal related  momenta $\bar{k}_1, -\bar{k}_1$ where the hybrid Wannier center $\nu=1/2$. The $\mathcal{C}_2 \Theta$ symmetry mandates that these unpinned degeneracies can be only moved [c.f. Appendix~\ref{sec:nestedPartial} and Ref.~\onlinecite{bra19}] pairwise (as required by time-reversal), and consequently cannot be destroyed. Hence, the winding of the Wilson loop spectrum is robust, which allows for the definition of a fragile topological phase in insulators with one occupied Kramers pair of bands. Furthermore, the Wilson loop winding can occur independent of the $\mathbb{Z}_2$ line invariants, thus suggesting that the complete classification in systems with a trivial FKM invariant is $\mathbb{Z}_2 \times \mathbb{Z}_2 \times \mathbb{Z}_2$, where the third ${\mathbb Z}_2$ invariant discriminates between gapped and winding Wilson loop spectra. 

\begin{figure}
\includegraphics[width=.98\columnwidth]{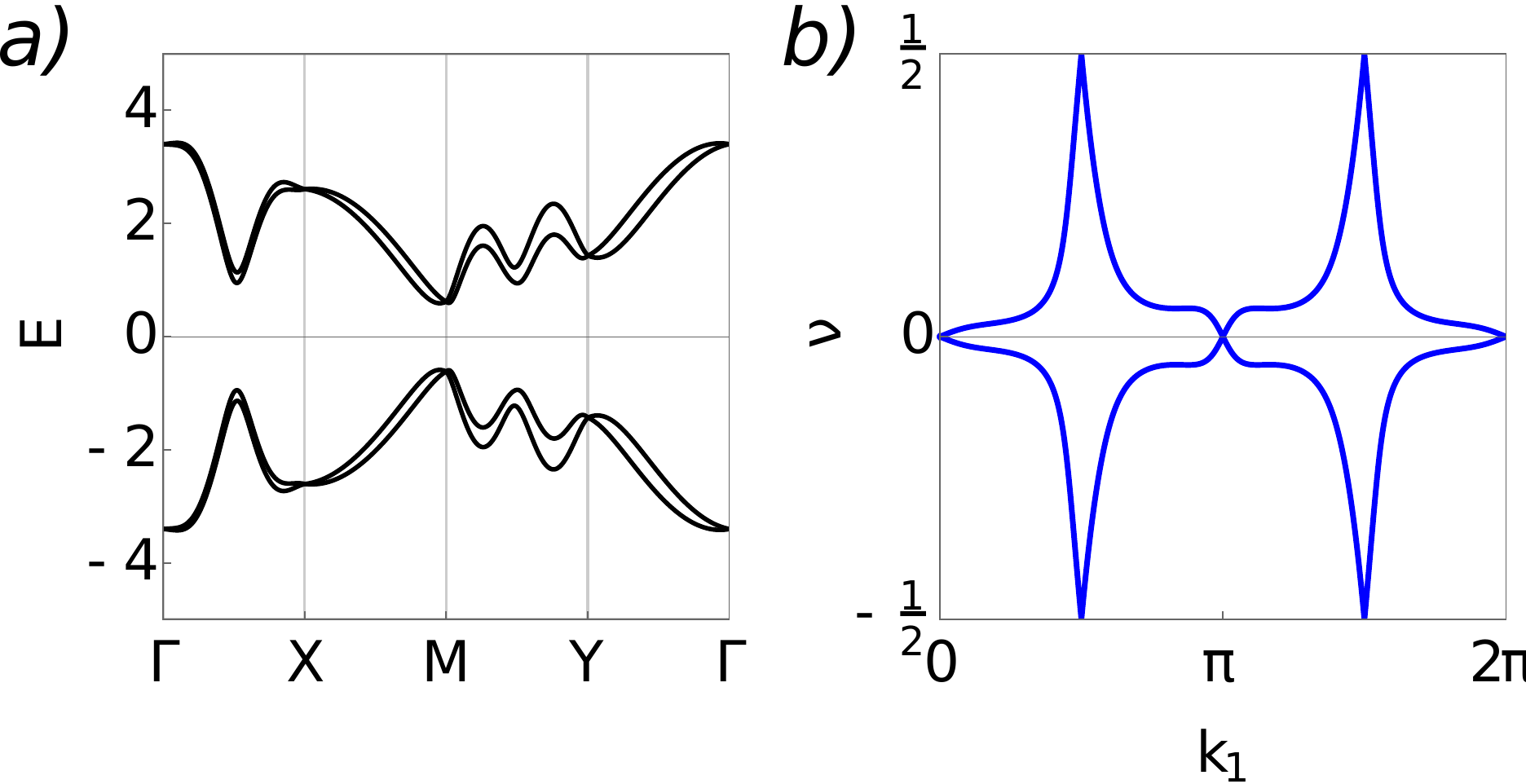}
\caption{(color online) (a) Band structure of the $N_F=2$ fragile topological insulator with twofold rotation and time-reversal symmetry. Energies have been measured in units of $t$. There are no degeneracies other than those required by time-reversal symmetry. (b) The Wilson loop spectrum along the $k_1$ direction for the half-filled insulating state. See Appendix \ref{sec:fragile2} for more details.}
\label{fig:fig2}
\end{figure}

To verify the existence of the fragile topological phase discussed above, we introduce a four-band tight-binding model on a ${\mathcal C}_3$ and mirror symmetry broken honeycomb lattice [see Appendix \ref{sec:fragile2} for the corresponding tight-binding model] with a full spectral gap at half-filling [see Fig.~\ref{fig:fig2}(a)]. It can be thought of as being made of two coupled Chern insulators with opposite Chern numbers ${\mathcal C}=\pm 2$, thereby respecting time-reversal symmetry. In Fig.~\ref{fig:fig2}(b) we show the Wilson loop spectrum along the $k_1$ direction, which displays the non-trivial winding discussed above. We close this section by emphasizing that the existence of the fragile topological phase does not strictly rely on the existence of a single PEBR. In Appendix \ref{sec:fragile2}, we introduce a ${\mathcal C}_4$ symmetric tight-binding model on the square lattice where the $N_F=2$ atomic insulating states can be generally represented in terms of symmetric WFs centered at the maximal Wyckoff positions $1a=\left\{0,0\right\}$ and $1b=\left\{1/2,1/2\right\}$, which possess
distinguishable PEBRs. The symmetry content of the occupied bands of our model is compatible with an atomic insulator with a Wannier Kramers pair centered at $1b$. However, inspection of the Wilson loop spectrum firmly establishes it as being a topological insulator of the fragile type. 

\section{$\mathbb{Z}_2 \times \mathbb{Z}_2 \times \mathbb{Z}_2$ classification with $N_F=4$: a new fragile topological phase}
With the $\mathbb{Z}_2 \times \mathbb{Z}_2 \times \mathbb{Z}_2$ classification of $N_F=2$ insulating phases in our hands, we next consider insulators with $N_F=4$. We will follow the same strategy used in the preceding section, and enumerate and classify all the existing atomic insulating phases. It is easy to see that there exist seven distinct insulating states representable in terms of symmetric WFs. In fact, with two Wannier Kramers pair in the system, their centers will either lie at two ${\mathcal C}_2$ related non-maximal Wyckoff positions or at two distinct maximal Wyckoff positions. Therefore, the two $\mathbb{Z}_2$ line invariants $\nu^{\mathcal M}(k_{1,2}=0)$ are insufficient to classify these  states. Now we will show, using a procedure similar to the ``nested" Wilson loop one of topological multipole insulators~\cite{ben17,fra18}, 
that it is possible to obtain an additional $\mathbb{Z}_2$ invariant by identifying two sectors in the Wilson loop spectrum, each of which carries its own topological content, {\it i.e.} its quantized partial polarization. 

\begin{figure}
\includegraphics[width=\columnwidth]{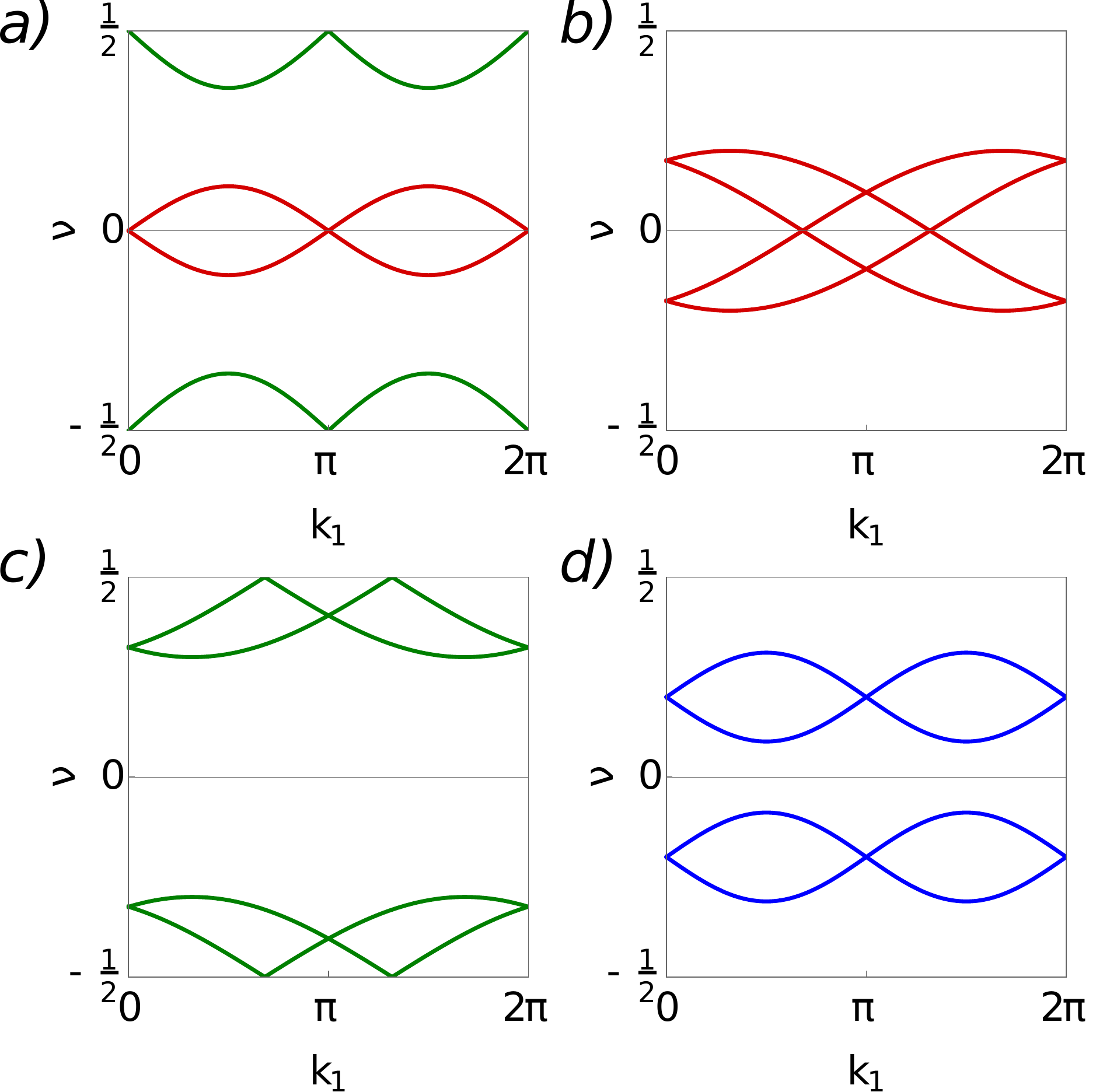}
\caption{(color online) Schematic drawings of the Wilson loop spectra for the $N_F=4$ atomic insulating states in the $p2$ wallapaper group. Panel (a) corresponds to four different atomic insulating states, where the pair of bands around $\nu=0$ ($\nu=1/2$) can have a Wannier center at $1a$ or $1b$ ($1c$ or $1d$), respectively, which can be determined by calculating their nested partial polarizations. Panel (b) corresponds to an atomic insulator with Wannier Kramers pairs centered at $1a \oplus 1b$, while panel (c) is for $1c \oplus 1d$. In panel (d) the Wannier functions are centered at ${\mathcal C}_2$ related generic points in the unit cell.}
\label{fig:fig3}
\end{figure}

We recall that 
the essential characteristic of a generic atomic insulating state is the presence of a Wannier gap in the Wilson loop spectrum. Its chiral symmetry, dictated by the ${\mathcal C}_2 \Theta$ symmetry, then 
allows us to distinguish two regions, one symmetrically centered around $\nu=0$ and one symmetrically centered around $\nu=1/2$, each possessing both twofold rotation and time-reversal symmetry, and populated by Kramers related pairs of Wannier bands. We have plotted the possible Wilson loop spectra for two Kramers pairs in Fig.~\ref{fig:fig3}, where the red bands are centered around $\nu=0$ and the green bands around $\nu=1/2$. The blue bands can be seen as centered around either point \footnote{Since we have to take a region symmetrically centered around $\nu=0,1/2$ we have to either include both or neither of the blue bands.}. Obviously, the parity of the pairs of Wannier bands belonging to the gapped region centered around $\nu=1/2$ can be linked to the line invariants $\nu^{\mathcal M}_{k_{1,2}=0}$. Considering for instance the spectrum of the Wilson loop $\mathcal{W}_{(k_1, k_2+ 2\pi) \leftarrow (k_1,k_2)}$ and further splitting the Wannier bands in two time-reversed channels, we immediately find that $\nu^{\mathcal M}_{k_1=0}=0$ ($\nu^{\mathcal M}_{k_1=0}=1$) if the Wilson loop spectrum region centered at $\nu=1/2$ is populated by an even [c.f. Figs.~3(b)-(d)]  (odd [c.f. Fig.~3(a)]) number of pairs of Wannier bands.
Furthermore, we can obtain two distinct $\mathbb{Z}_2$ invariants for the two disconnected regions of the $k_1$ dependent Wilson loop spectrum as follows. Let us consider the Wilson loop operator $\mathcal{W}_{(k_1, k_2+ 2\pi) \leftarrow (k_1,k_2)}$, choosing its base point on the time-reversal and twofold rotation symmetric line $k_2=0$ [c.f. Fig.~\ref{fig:bz}]. The corresponding eigenstates $\ket{\nu^j_{e_2; (k_1,0)}}$, where the subscript $e_2$ specifies the $k_2$ direction of the Wilson loop, satisfy 
$$\mathcal{W}_{(k_1, 2 \pi ) \leftarrow (k_1,0)} \ket{\nu^j_{e_2; (k_1,0)}} =  e^{2 \pi i \nu_j(k_1)}  \ket{\nu^j_{e_2; (k_1,0)}}, $$
and allow us to define the Wannier basis \cite{ben17,ben17-2},
$\ket{w^j_{e_2; (k_1,0)}}=\sum_n \ket{u^n_{(k_1,0)}} \left[\nu^j_{e_2; (k_1,0)} \right]^n $, where $n=1,\ldots, N_F$.  
Since the quantized partial polarization  associated to the Bloch Hamiltonian eigenfunctions $\ket{u^n_{(k_1,0)}}$ is unchanged by a general $U(N_{F})$  transformation, it follows that the $\mathbb{Z}_2$ invariant $\nu^{\mathcal M}_{k_2 =0}$ can be equivalently computed in the Wannier band eigenbasis $\ket{w^j_{e_2; (k_1,0)}}$. More importantly, working in such a basis allows us to decompose $\nu^{\mathcal M}_{k_2 =0}$ into two different $\mathbb{Z}_2$ invariants, which we dub as $\nu^{\mathcal M ; 0}_{k_{2}=0}$ and $\nu^{\mathcal M ; 1/2}_{k_{2}=0}$, corresponding to the ``nested" quantized partial polarizations for the two gapped sectors of the Wilson loop spectrum (the red and green bands in Fig.~\ref{fig:fig3}, respectively). 
This is because, as mentioned above, the two gapped regions separately satisfy both time-reversal and twofold rotation symmetry, which guarantees that the partial polarization of the corresponding Wannier band eigenstates is quantized. Note that Wannier bands only respect twofold rotation and time-reversal symmetry when the Wilson loop base points lie on a mirror symmetric line.

Having obtained three distinct $\mathbb{Z}_2$ topological invariants, we can now classify the atomic insulating phases enumerated above.  Fig.~\ref{fig:fig3}(a) schematically shows the $k_1$-dependent Wilson loop spectrum when the two gapped sectors are each populated with one pair of Wannier bands, and thus $\nu^{\mathcal M}_{k_1=0}=1$. The gapped sector centered around $\nu=0$ is further characterized by the ${\mathbb Z}_2$ invariant $\nu^{\mathcal M ; 0}_{k_{2}=0}$, and its value dictates whether the Wannier Kramers pair is centered at the maximal Wyckoff position $1a$ ($\nu^{\mathcal M ; 0}_{k_{2}=0}=0$) or $1b$ ($\nu^{\mathcal M ; 0}_{k_{2}=0}=1$). The same argument 
can be applied to the gapped sector centered at $\nu=1/2$ to set apart Wannier Kramers pairs centered at $1c$ ($\nu^{\mathcal M ; 1/2}_{k_{2}=0}=0$) and $1d$ ($\nu^{\mathcal M ; 1/2}_{k_{2}=0}=1$). This, in turn, allows us to catalogue four distinct atomic insulating states. 

\begin{table}[t!]
\begin{tabular}{|c|c|c|c|}
\hline 
Wyckoff positions & $\nu^{\mathcal M}_{k_1=0}$ & $\nu^{\mathcal M;0}_{k_2=0}$ & $\nu^{\mathcal M;1/2}_{k_2=0}$  \tabularnewline
\hline 
$1a \oplus 1c$ & 1 & 0 & 0  \tabularnewline 
\hline
$1a \oplus 1d$ & 1 & 0 & 1 \tabularnewline 
\hline 
$1b \oplus 1c $ & 1 & 1 & 0 \tabularnewline 
\hline
$1b \oplus 1d$ & 1 & 1 & 1 \tabularnewline
\hline
$1a \oplus 1b$ & 0 & 1 & 0 \tabularnewline
\hline 
$1c \oplus 1d$ & 0 & 0 & 1 \tabularnewline 
\hline 
$\boldsymbol{\nu} \oplus -\boldsymbol{\nu} $ & 0 & 0 & 0 \tabularnewline
\hline
\end{tabular}
\caption{The classification of atomic insulating states in the $p2$ wallpaper group when two occupied Kramers pairs of bands are occupied, {\it i.e.} $N_F=4$. The first column indicates the centers of charge of the Wannier Kramers pairs; the second column is the ${\mathbb Z}_2$ line invariant of the full Wilson loop spectrum; the second and third columns are the invariants derived from the nested Wilson loops, which obey the sum rule $\left(\nu^{\mathcal M ; 0}_{k_{2}=0} + \nu^{\mathcal M ; 1/2}_{k_{2}=0}\right)~\textrm{mod}~2=\nu^{\mathcal M}_{k_{2}=0}$. The last row refers to insulators where the Wannier Kramers pairs are centered at ${\mathcal C}_2$ related non-maximal Wyckoff positions.}
\label{tab:tabNf4} 
\end{table}

Next, we consider insulating states where the Wannier bands occupy only one gapped sector of the Wilson loop spectrum, and thus $\nu^{\mathcal M}_{k_1=0}=0$. Fig.~\ref{fig:fig3}(b),(c),(d) show the allowed possibilities for the Wannier bands. They can either realize a connected pair with two protected degeneracies at time-reversal related momenta $(\bar{k}_1, -\bar{k}_1)$  
or can come in disconnected pairs, in which case the two pairs can be arbitrarily assigned to the $\nu=0$ or the $\nu=1/2$ sector. Let us first inspect the value the invariants $\nu^{\mathcal M ; 0}_{k_{2}=0}$ ($\nu^{\mathcal M ; 1/2}_{k_{2}=0}$) assume for the connected pair of Wannier bands shown in Fig.~\ref{fig:fig3}(b),(c). We can divide the four Wannier bands in two time-reversed channels, that each possess ${\mathcal C}_2 \Theta$ symmetry. Then, an essential twofold degeneracy in one channel at $\nu=0$ ($\nu=1/2$)  implies a $\pi$ Berry phase [see Appendix~\ref{sec:nestedPartial} and Ref.~\onlinecite{ahn18}], and consequently the nested line invariant  $\nu^{\mathcal M ; 0}_{k_{2}=0}$ ($\nu^{\mathcal M ; 1/2}_{k_{2}=0}$) is enforced to be $1$. As a result, the schematic Wannier bands shown in Fig.~\ref{fig:fig3}(b),(c)  correspond to 
the atomic insulating phase with Wannier Kramers pairs centered at $1a \oplus 1b$ and $1c \oplus 1d$ respectively. 
Using similar arguments, we also find that the disconnected Wannier bands of Fig.~\ref{fig:fig3}(d) are characterized by a zero nested partial polarization [see Appendix \ref{sec:nestedPartial}]. Therefore, in this atomic insulating state the Wannier Kramers pairs are centered at two non-maximal Wyckoff positions related to each other by the twofold rotation symmetry.
All in all, we have thus reached the classification summarized in Table~\ref{tab:tabNf4} of the seven distinct atomic insulating states realizable in the $p2$ wallpaper group with four occupied bands.  

When comparing this with the eight allowed configurations for the three ${\mathbb Z}_2$ invariants, one can immediately recognize that an insulating state characterized by the two nested quantization polarization invariants $\nu^{\mathcal M;0}_{k_2=0}=\nu^{\mathcal M;1/2}_{k_2=0}=1$ with $\nu^{\mathcal M;0}_{k_1=0}=0$ cannot be represented in terms of symmetric exponentially localized Wannier functions. In fact, such a configuration featuring essential degeneracies at unpinned momenta $k_1$ both around $\nu=0$ and $\nu=1/2$ would necessarily imply the closing of the Wannier gap and hence a non-trivial winding of the Wilson loop. We thus conclude that such an insulator corresponds to a topologically non-trivial phase of the fragile type. Its stability against symmetry-allowed perturbations is rooted in the fact that the possible local annihilation of the degeneracies on the $\nu=0$ or $\nu=1/2$ line requires a change of the line invariant $\nu^{\mathcal M}_{k_2=0}=\left(\nu^{\mathcal M ; 0}_{k_{2}=0} + \nu^{\mathcal M ; 1/2}_{k_{2}=0}\right)~\textrm{mod}~2$, which is only possible with a bandgap closing-reopening point.

\begin{figure}
\includegraphics[width=\columnwidth]{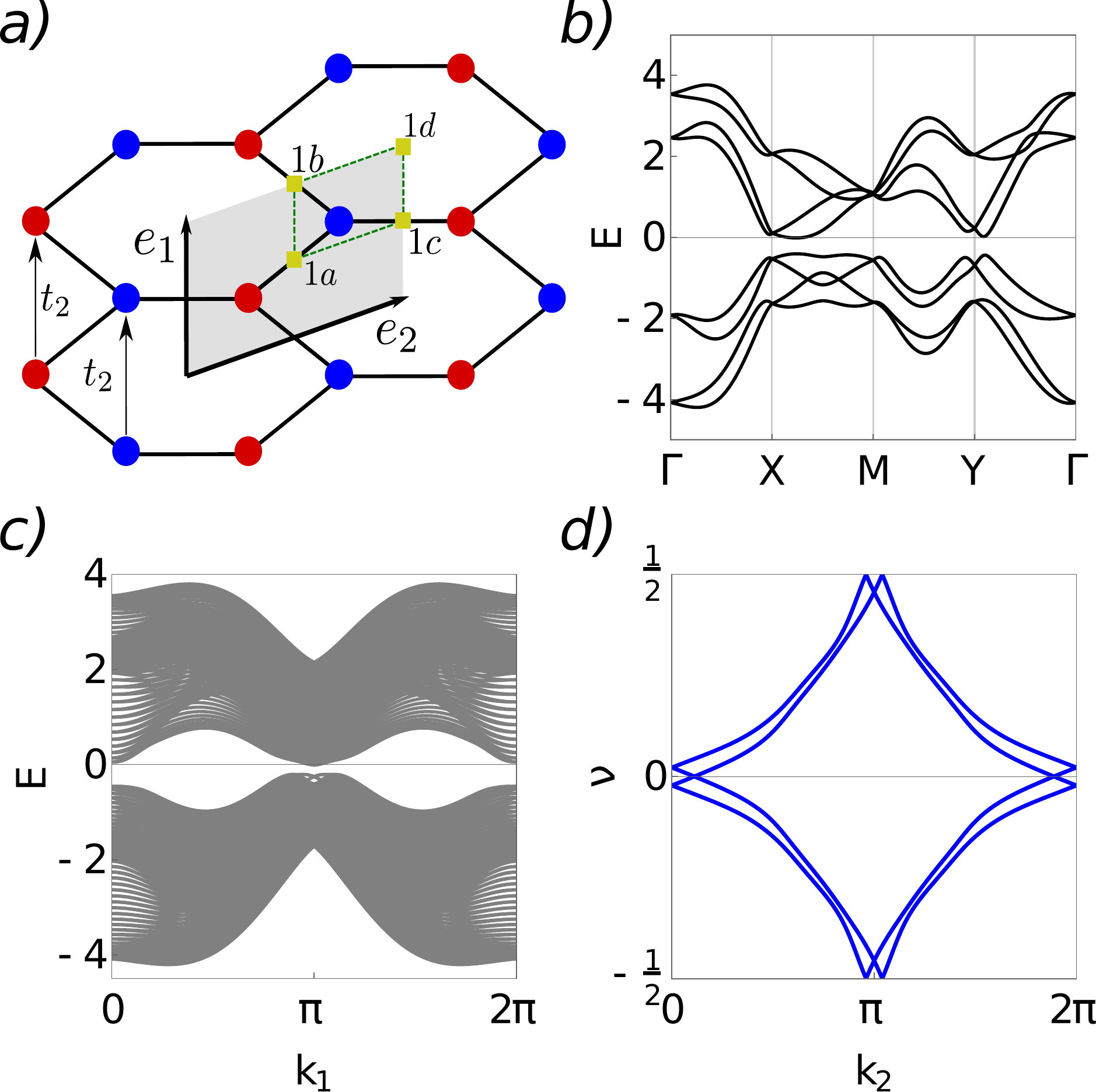}
\caption{(color online) (a) Top view of the strained honeycomb bilayer realizing the $N_F=4$ fragile topological phase. The intralayer spin-dependent hopping amplitude $t_2$ has been taken only along the zigzag direction to amplify the threefold rotation symmetry breaking. (b) Bulk bands showing a full spectral gap at half-filling. The parameter set is specified in Appendix \ref{sec:fragile4}. (c) The corresponding spectrum in a ribbon geometry demonstrate the insulating nature of the edges. (d) Wannier bands along the $k_2$ direction. The Wilson loop in the $k_1$ direction also show a similar winding. }
\label{fig:fig4}
\end{figure}

Let us now present a model realization of this novel fragile topological insulating phase. The model is built by stacking two quantum spin-Hall insulators 
on the honeycomb lattice -- the so-called Kane-Mele model~\cite{kan05b} -- with opposite sign of the spin-dependent next-nearest neighbor hopping $t_2$ parametrizing the spin-orbit coupling strength. Inversion symmetry is explicitly broken by considering a chemical potential difference between the two layers while the threefold rotation symmetry breaking due to, {\it e.g.}, a uniaxial strain [c.f. Fig.~\ref{fig:fig4}(a)] is incorporated taking direction dependent hopping amplitudes $t_2$. We also break the ${\mathcal M}_z$ symmetry by introducing a Rashba spin-orbit coupling term. Being composed of two quantum spin Hall insulators, the FKM invariant of the half-filled model is trivial, and 
with an explicit interlayer coupling
the helical edge states disappear [see Appendix \ref{sec:fragile4} for the model Hamiltonian and Fig.~\ref{fig:fig4}(b) for the ribbon spectrum]. A direct computation of the Wilson loop spectrum [c.f. Fig.~\ref{fig:fig4}(d)] shows the non-trivial winding with the line invariants $\nu^{\mathcal M}_{k_{1,2}=0}=0$ that present an obstruction to the Wannier representation of this phase. In Appendix \ref{sec:fragile4}, we also present a spinful model inspired by the $p_{x,y}$ orbital model presented in Ref.~\cite{can18} that also realizes the $N_F=4$ fragile topological phase discussed above. 

\begin{table}
\begin{tabular}{|c|c|c|c|}
\hline 
Wyckoff positions & $\nu_{k_{1}=0}^{\mathcal{M}}$ & $\nu_{k_{2}=0}^{\mathcal{M};0}$ & $\nu_{k_{2}=0}^{\mathcal{M};1/2}$\tabularnewline
\hline 
\hline 
$1a\oplus1b\oplus1c$ & $1$ & $1$ & 0\tabularnewline
\hline 
$1a\oplus1b\oplus1d$ & $1$ & $1$ & 1\tabularnewline
\hline 
$1a\oplus1c\oplus1d$ & $0$ & $0$ & 1\tabularnewline
\hline 
$1b\oplus1c\oplus1d$ & $0$ & $1$ & 1\tabularnewline
\hline 
$1a\oplus\boldsymbol{\nu}\oplus-\boldsymbol{\nu}$ & $0$ & $0$ & 0\tabularnewline
\hline 
$1b\oplus\boldsymbol{\nu}\oplus-\boldsymbol{\nu}$ & $0$ & $1$ & 0\tabularnewline
\hline 
$1c\oplus\boldsymbol{\nu}\oplus-\boldsymbol{\nu}$ & $1$ & $0$ & 0\tabularnewline
\hline 
$1d\oplus\boldsymbol{\nu}\oplus-\boldsymbol{\nu}$ & $1$ & $0$ & 1\tabularnewline
\hline 
\end{tabular}
\caption{The $\mathbb{Z}_2 \times \mathbb{Z}_2 \times \mathbb{Z}_2$ classification of atomic insulating states in the $p2$ wallpaper group when three occupied Kramers pairs of bands are occupied, {\it i.e.} $N_F=6$, indicating the relation between the Wannier Kramers pairs center of charges and the (``nested") quantized partial polarization topological invariants. This classification is generically valid for an arbitrary number of occupied bands $N_F=4 n +2$ with the integer $n \geq 1$, which will only include more unpinned pairs of Kramers pairs. 
}
\label{tab:tabNf6}
\end{table}

\section{More occupied bands} 
Contrary to the $N_F=2$ topologically non-trivial phase, which is trivialized only when certain Kramers pairs of bands are added , the $N_F=4$ topological insulator discussed above is intensively fragile: it is trivialized by the addition of a generic Kramers pair of bands. This assertion can be immediately proved noticing that for a generic $N_F=6$ insulating state, the ${\mathbb Z}_2 \times {\mathbb Z}_2 \times {\mathbb Z}_2$ classification is saturated by enumerating the phases with symmetric Wannier function. In fact, with three Wannier Kramers pairs in the system, their centers can either lie on three distinct maximal Wyckoff positions, or two Wannier pairs sit at ${\mathcal C}_2$-related non-maximal Wyckoff position with a third pair located at one maximal Wyckoff position. Inspecting the possible features of the Wilson loop spectrum and iterating the arguments presented in the former sections we reach the classification summarized in Table~\ref{tab:tabNf6}. Note that this classification is generally valid for $N_F=4 n+2$ and $n \geq 1$. In fact, by adding two Wannier Kramers pairs to a state with $N_{F}=6$, we will end up in one of the $N_{F}=6$ configurations [c.f. Table \ref{tab:tabNf6}] with the addition of two Wannier Kramers pair centered at unpinned two-fold rotation symmetric momenta, which do not change the $\mathbb{Z}_2$ invariants.

Finally, in Table~\ref{tab:tabNf8}  we also provide the classification of atomic insulators with four Wannier Kramers, which is also valid for a generic number of occupied bands $N_F=4 n$ and $n>1$. Note that the distribution of ${\mathbb Z}_2$ invariants is strictly equivalent to the case of four occupied bands. However, the topological non-trivial fragile phase is substituted by an atomic insulator where the four Wannier Kramers pairs are centered at the four maximal Wyckoff positions. In this configuration, in fact, the Wilson loop spectrum is the superposition of Fig.~\ref{fig:fig3}(b) and Fig.~\ref{fig:fig3}(c) which is allowed with a full Wannier gap with a minimum number of eight Wannier bands.

 \begin{table}
 \begin{tabular}{|c|c|c|c|}
\hline 
Wyckoff positions & $\nu_{k_{1}=0}^{\mathcal{M}}$ & $\nu_{k_{2}=0}^{\mathcal{M};0}$ & $\nu_{k_{2}=0}^{\mathcal{M};1/2}$\tabularnewline
\hline 
\hline 
$1a\oplus1b\oplus1c\oplus1d$ & $0$ & $1$ & $1$\tabularnewline
\hline 
$\boldsymbol{\nu}_{1}\oplus-\boldsymbol{\nu}_{1}\oplus\boldsymbol{\nu}_{2}\oplus-\boldsymbol{\nu}_{2}$ & $0$ & $0$ & $0$\tabularnewline
\hline 
$1a\oplus1b\oplus\boldsymbol{\nu}\oplus-\boldsymbol{\nu}$ & $0$ & $1$ & $0$\tabularnewline
\hline 
$1a\oplus1c\oplus\boldsymbol{\nu}\oplus-\boldsymbol{\nu}$ & $1$ & $0$ & $0$\tabularnewline
\hline 
$1a\oplus1d\oplus\boldsymbol{\nu}\oplus-\boldsymbol{\nu}$ & $1$ & $0$ & $1$\tabularnewline
\hline 
$1b\oplus1c\oplus\boldsymbol{\nu}\oplus-\boldsymbol{\nu}$ & $1$ & $1$ & $0$\tabularnewline
\hline 
$1b\oplus1d\oplus\boldsymbol{\nu}\oplus-\boldsymbol{\nu}$ & $1$ & $1$ & $1$\tabularnewline
\hline 
$1c\oplus1d\oplus\boldsymbol{\nu}\oplus-\boldsymbol{\nu}$ & $0$ & $0$ & $0$\tabularnewline
\hline 
\end{tabular}
\caption{The $\mathbb{Z}_2 \times \mathbb{Z}_2 \times \mathbb{Z}_2$ classification of atomic insulating states in the $p2$ wallpaper group when four occupied Kramers pairs of bands are occupied, {\it i.e.} $N_F=8$, indicating the relation between the Wannier Kramers pairs center of charges and the (``nested") quantized partial polarization topological invariants. This classification is generically valid for an arbitrary number of occupied bands $N_F=4 n $ with the integer $n > 1$, which will only include more unpinned pairs of Kramers pairs. 
}
\label{tab:tabNf8}
\end{table}

\section{Conclusions} 
 In this paper, we presented a classification of gapped insulating phases that cannot be diagnosed using crystalline symmetry eigenvalues. We have showcased two-dimensional crystals in the wallpaper group $p2$ where all gapped phases have the same physical elementary band representation, but they can be nevertheless classified with three ${\mathbb Z}_2$ topological invariants: the quantized nested partial polarizations -- partial Berry phases -- along high-symmetry lines in the two-dimensional Brillouin zone of the system. 
 
 Using the ensuing ${\mathbb Z}_2 \times {\mathbb Z}_2 \times {\mathbb Z}_2$ classification, we have been able to classify all atomic insulating states and identify non-Wannierazible topological crystalline phases protected by twofold rotation symmetry and time-reversal symmetry. Since the crystal does not possess boundaries that are left invariant under the protecting twofold rotation symmetry, these topological phases do not display gapless anomalous boundary modes although their bulk is topologically non-trivial. Instead, they represent an example of the recently discovered fragile topology, and thus they can be trivialized with the addition of atomic valence bands. In this respect, we wish to emphasize that the fragile topological phase realized with two occupied valence bands, which is similar in nature to the fragile phases recently discussed in the literature in other wallpaper groups 
  does not necessarily decay into a Wannierazible atomic insulating state when an additional Kramers related pair of bands are introduced. In fact, such band addition might lead to our novel $N_F=4$ topological crystalline phase whose Wilson loop winding 
is strictly protected by the quantization of the nested quantized partial Berry phase in the presence
of time-reversal and twofold rotation symmetries. 
 
 An interesting direction for future research is the extension of the classification presented here to other wallpaper and space groups where the symmetry data of the valence bands could be combined with Berry phase invariants to search for new topological electronic materials. Furthermore, the Berry phase invariants for atomic insulating phases can be also exploited to obtain, using the Wannier centers flow of hybrid Wannier functions~\cite{son17,mie18}, topological invariants for higher-order topological insulators with helical hinge modes in non-centrosymmetric crystals.  
 
\begin{acknowledgments}
C.O. acknowledges support from a VIDI grant (Project 680-47-543) financed by the Netherlands Organization for Scientific Research (NWO). This work is part of the research programme of the Foundation for Fundamental Research on Matter (FOM), which is part of the Netherlands Organization for Scientific Research (NWO). S.K. acknowledges support from a NWO-Graduate Program grant. 
\end{acknowledgments}

 \appendix
 \section{Spin-orbit coupled Aubry-Andr\'e-Harper model}\label{sec:harper} 
To analyze 1D atomic chains with time-reversal and mirror symmetry with respect to a mirror point, we consider a tight-binding model~\cite{lau16}
for spin-1/2 electrons corresponding to a generalized Aubry-Andr\'e-Harper model~\cite{har55,aub80,gan13} 
\begin{eqnarray*}
\mathcal{H} & =&\sum_{j,\sigma}\left[t_{0}+\delta t\cos\left(\pi j+\phi_{t}\right)\right]c_{j+1,\sigma}^{\dagger}c_{j,\sigma} \\
& & +i\sum_{j,\sigma,\sigma'}\left[\lambda_{0}+\delta\lambda\cos\left(\pi j+\phi_{\lambda}\right)\right]c_{j+1,\sigma}^{\dagger}s_{\sigma\sigma'}^{y}c_{j,\sigma'}\\ 
 && +\sum_{j,\sigma}\left[V_{0}+\delta V\cos\left(j\pi/2+\phi_{V}\right)\right]c_{j,\sigma}^{\dagger}c_{j,\sigma}+\mathrm{H.c.},
\end{eqnarray*}
where $c_{j,\sigma}^{\dagger}$
is the creation operator for an electron at site $j$ with spin $\sigma$ ($\sigma=\uparrow,\downarrow$), and $s^i$ are the conventional Pauli matrices. The Hamiltonian contains harmonically modulated nearest-neighbor hopping, spin-orbit coupling and onsite potentials of amplitudes $\delta t$, $\delta\lambda$, and $\delta V$, and phases $\phi_t$, $\phi_{\lambda}$ and $\phi_V$. The periodicities of the modulated hopping and spin-orbit coupling have been chosen to be of two lattice sites while the periodicity of the onsite potential is four lattice sites. Moreover, $t_{0},\lambda_{0}$ and $V_{0}$ are the site-independent amplitudes of the hopping,
spin-orbit coupling and on-site potential.  The model possesses time-reversal symmetry whereas mirror symmetry is preserved only for specific values of the phases $\phi_{t,\lambda,V}$. In Fig.~\ref{fig:fig1} we have chosen the parameter set $\phi_{t}=\phi_{\lambda}=\pi$, $\lambda_{0}=0.5t_{0}$, $\delta\lambda_{0}=-0.3 t_0$ and 
$\delta V=t_{0}$. The Wilson loop eigenvalues shown in Fig.~\ref{fig:fig1}(a) have been obtain using the mirror-symmetric value $\phi_V=-\pi/4$ while sweeping the dimerized hopping amplitude $\delta t$. In Fig.\ref{fig:fig1}(b), instead, we have fixed $\delta t=-0.25 t_0$ while sweeping $\phi_V$ away from the mirror symmetric point. 

\section{Nested partial polarization in Wilson loop spectra with $\mathcal{C}_{2}$
and $\Theta$ symmetry}\label{sec:nestedPartial}

Here we show that the nested partial polarizations are well-defined
quantities in gapped Wilson loop spectra, and that they are quantized
in the presence of $\mathcal{C}_{2}$ and $\Theta$ symmetry. In addition, we 
calculate the partial polarizations for various Wilson loop spectra.

Let us start by examining how the symmetries act on the Wilson loop.
Consider the Wilson loop operator $\mathcal{W}_{\left(k_{1},2\pi\right)\leftarrow\left(k_{1},0\right)}$,
$\mathcal{C}_{2}$ and $\Theta$ symmetry then require \cite{bra19,ale14}
\begin{align*}
\mathcal{C}_{2}\mathcal{W}_{\left(k_{1},2\pi\right)\leftarrow\left(k_{1},0\right)}\mathcal{C}_{2}^{\dagger} & =\mathcal{W}_{\left(-k_{1},2\pi\right)\leftarrow\left(-k_{1},0\right)}^{\dagger},\\
\Theta\mathcal{W}_{\left(k_{1},2\pi\right)\leftarrow\left(k_{1},0\right)}\Theta^{\dagger} & =\mathcal{W}_{\left(-k_{1},2\pi\right)\leftarrow\left(-k_{1},0\right)}^{\dagger},
\end{align*}
where the complex conjugate on the right-hand side comes from the
fact that both symmetries send $k\rightarrow-k$ and hence reverse
the contour of the Wilson loop operator. Furthermore, $\mathcal{C}_{2}$
relates the eigenvalues of the Wilson loop operator
\begin{align*}
\left\{ \nu_{i}\left(k_{1}\right)\right\}  & =\left\{ -\nu_{i}\left(-k_{1}\right)\right\} ,
\end{align*}
and time-reversal relates 
\begin{align*}
\left\{ \nu_{i}\left(k_{1}\right)\right\}  & =\left\{ \nu_{i}\left(-k_{1}\right)\right\} ,
\end{align*}
where $\left\{ \right\} $ denotes the set of eigenvalues. Hence $\mathcal{C}_{2}\Theta$
enforces a chiral symmetry in the Wilson loop spectrum.

Now let us show, following Ref.~\cite{bra19}, that
a single crossing in the Wilson loop spectrum is locally protected
by the combination of $\mathcal{C}_{2}$ and $\Theta$ symmetry. Let us work in a
basis where $\mathcal{C}_{2}\Theta=\mathcal{K},$ where $\mathcal{K}$ indicates complex conjugation. The symmetry restriction
on the Wilson loop operator is then
\begin{align*}
\mathcal{K}\mathcal{W}_{\left(k_{1},2\pi\right)\leftarrow\left(k_{1},0\right)}\mathcal{K} & =\mathcal{W}_{\left(k_{1},2\pi\right)\leftarrow\left(k_{1},0\right)},
\end{align*}
since $\mathcal{C}_{2}\Theta$ sends $k\rightarrow k$. Since the Wilson loop operator in this basis
is an $SO\left(N\right)$ matrix, we can write it as the exponential
of an Hermitian matrix $\mathcal{H_{W}}$, which is restricted by
$\mathcal{C}_{2}\Theta$ such that 
\begin{align*}
\mathcal{H_{W}}\left(k_{1}\right) & =-\mathcal{H_{W}}\left(k_{1}\right)^{*}.
\end{align*}
Near a two-band crossing, this restriction means that locally $\mathcal{H_{W}}\left(k_{1}\right)=k_{1}\cdot\sigma_{y}$. 
A single twofold degeneracy on the $\nu=0,1/2$ lines
cannot be gapped out without breaking $\mathcal{C}_{2}\Theta$ symmetry, but only moved on the line. Therefore, as for a Weyl point, the degeneracy can be only removed by pair annihilation.

\begin{figure}
\includegraphics[width=\columnwidth]{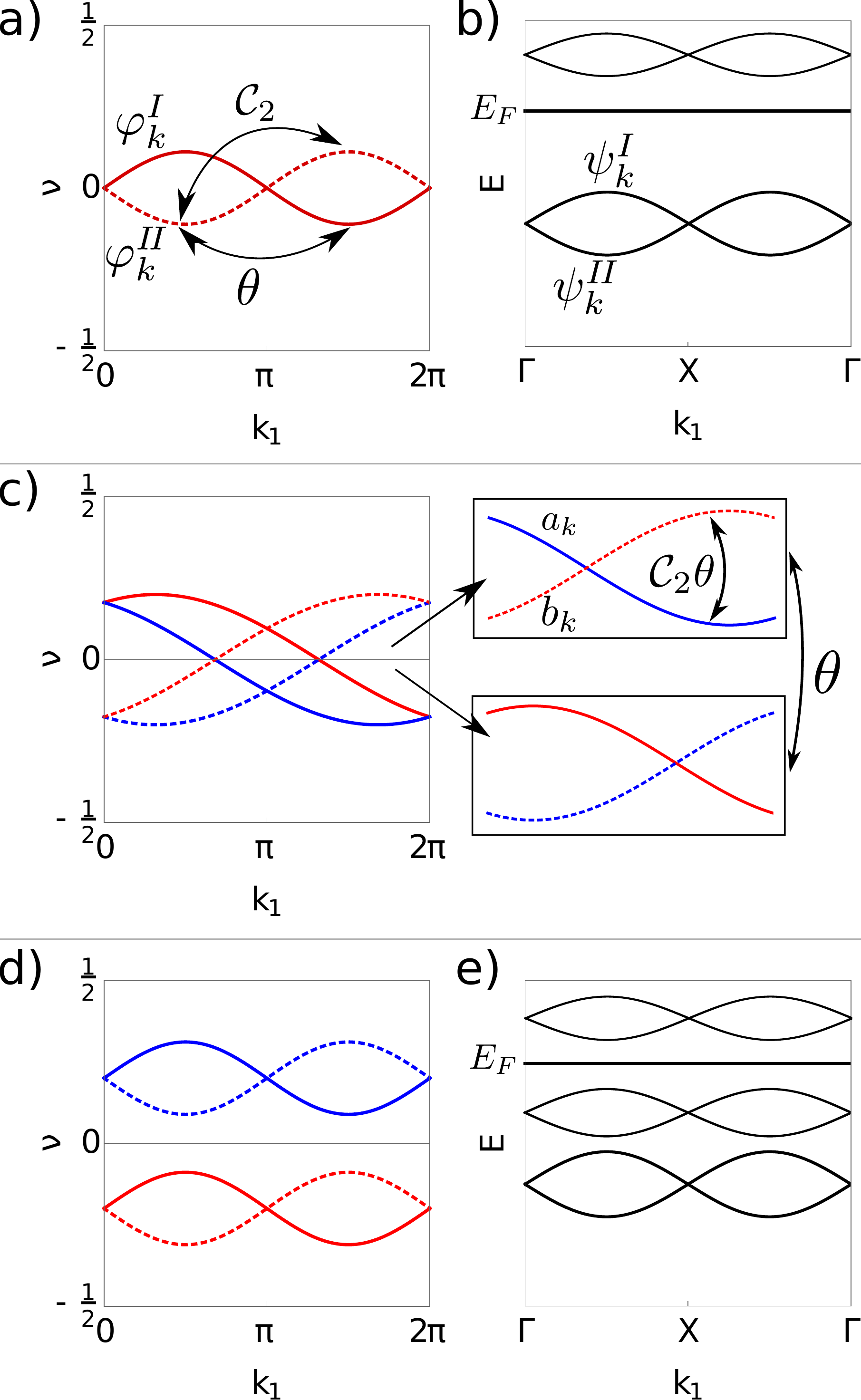}
\caption{(a) Schematic drawing of a generic Wilson loop spectrum with one occupied Kramers pair. The two Wannier bands $\varphi_{k}^{I}$ and $\varphi_{k}^{II}$ are related to each other by time-reversal symmetry, and are themselves $\mathcal{C}_{2}$ symmetric. (b) Generic band structure corresponding to (a), the Wannier states are obtained by linear combinations of the eigenstates of the Hamiltonian $\psi_{k}^{I}$,$\psi_{k}^{II}$. (c) Wilson loop spectrum for two occupied Kramers pairs with two crossings at $\nu=0$. The colors denote the two different Kramers pairs, and the $\mathcal{C}_{2}$ partners have a solid (dotted) line. The two time-reversal channels are depicted on the right, which are by themselves $\mathcal{C}_{2}\Theta$ symmetric. (d) Disconnected Wilson loop of two occupied Kramers pairs, again color denotes Krames pairs, dotted (solid) the $\mathcal{C}_{2}$ partners. (e) Corresponding band structure with two occupied Kramers pairs. \label{fig:wlsbz}}
\end{figure}

We now turn to the various possible Wilson loop spectra, and compute
their partial polarizations. In Fig. \ref{fig:wlsbz}(a) we have drawn
a generic Wilson loop spectrum for one occupied Kramers pair. The
corresponding Wannier bands are given by 
\begin{align*}
\varphi_{k}^{I} & =\alpha\psi_{k}^{I}+\beta\psi_{k}^{II},\\
\varphi_{k}^{II} & =\gamma\psi_{k}^{I}+\delta\psi_{k}^{II},
\end{align*}
where $\psi_{k}^{I}$ and $\psi_{k}^{II}$ are the Bloch waves (schematically
drawn in Fig. \ref{fig:wlsbz}(b) along the same contour), and the
coefficients are given by the eigenvectors of the Wilson loop matrix
[see also Sec. IV]. The Wannier bands in Fig. \ref{fig:wlsbz}(a)
are thus obtained by a unitary transformation on the occupied eigenstates
of the Hamiltonian [(]Fig. \ref{fig:wlsbz}(b)], and will be linear
combinations thereof. These Wannier bands satisfy [(]see again Fig.
\ref{fig:wlsbz}(a)],
\begin{align*}
\varphi_{k}^{I} & =e^{i\theta\left(k\right)}\mathcal{C}_{2}\varphi_{-k}^{I},\\
\varphi_{k}^{I} & =e^{i\phi\left(k\right)}\Theta\varphi_{-k}^{II}.
\end{align*}
The partial polarization is in this case given by the Berry phase
of $\varphi_{k}^{I}$. Since $\varphi_{k}^{I}$ is $\mathcal{C}_{2}$
symmetric, its Berry phase, and hence the partial polarization is
quantized to $0,\pi$.

Now consider two occupied Kramers pairs with two crossings at $\nu=0$
(Fig. \ref{fig:wlsbz}(c)). The colors indicate the Kramers partners,
and the dotted (solid) lines are $\mathcal{C}_{2}$ partners. To find
the partial polarization we split the bands into two time-reversal
channels. The only possibility that leaves us with periodic subsets
of bands is taking the solid blue and dotted red bands together, and
the solid red and dotted blue bands together (shown in Fig. \ref{fig:wlsbz}(c)
on the right).

Let us denote the solid blue Wannier band by $a\left(k\right)$, and
define the red dotted band $b\left(k\right)$ by 
\begin{align*}
b\left(k\right) & :=\mathcal{C}_{2}\theta\left(k\right).
\end{align*}
Clearly the
bands are not periodic, and we have 
\begin{align*}
a\left(2\pi\right) & =b\left(0\right),\\
b\left(2\pi\right) & =a\left(0\right).
\end{align*}
We now try to construct a periodic gauge by a basis transformation,
under which the partial polarization is invariant. We define 
\begin{align*}
\tilde{a}\left(k\right) & =\left[a\left(k\right)+b\left(k\right)\right]/2,\\
\tilde{b}\left(k\right) & =\left[a\left(k\right)-b\left(k\right)\right]/2.
\end{align*}
Now
\begin{align*}
\tilde{a}\left(2\pi\right) & =\left[a\left(2\pi\right)+b\left(2\pi\right)\right]/2\\
 & =\left[b\left(0\right)+a\left(0\right)\right]/2\\
 & =\tilde{a}\left(0\right),
\end{align*}
hence $\tilde{a}\left(k\right)$ is periodic, however
\begin{align*}
\tilde{b}\left(2\pi\right) & =\left[a\left(2\pi\right)-b\left(2\pi\right)\right]/2\\
 & =\left[b\left(0\right)-a\left(0\right)\right]/2\\
 & =-\tilde{b}\left(0\right),
\end{align*}
is anti-periodic. Multiplying by a phase and defining
\begin{align*}
\tilde{\tilde{b}}\left(k\right) & =e^{ik/2}\tilde{b}\left(k\right),
\end{align*}
remedies this situation. Under $\mathcal{C}_{2}\Theta$ we now have
\begin{align*}
\mathcal{C}_{2}\Theta\,\tilde{a}\left(k\right) & =\tilde{a}\left(k\right),\\
\mathcal{C}_{2}\Theta\,\tilde{\tilde{b}}\left(k\right) & =e^{-ik/2}\tilde{b}\left(k\right)\\
 & =e^{-ik}\tilde{\tilde{b}}\left(k\right).
\end{align*}
Using this, we can calculate the Berry phase of the two bands separately,
\begin{align*}
\gamma_{a} & =\int dk\,\tilde{a}\left(k\right)^{\dagger}i\partial_{k}\tilde{a}\left(k\right)\\
 & =\int dk\,\tilde{a}\left(k\right)^{\dagger}\left(\mathcal{C}_{2}\theta\right)^{\dagger}\left(\mathcal{C}_{2}\theta\right)i\partial_{k}\tilde{a}\left(k\right)\\
 & =-\int dk\,\tilde{a}\left(k\right)^{\dagger}i\partial_{k}\tilde{a}\left(k\right)\\
 & =-\gamma_{a}
\end{align*}
and 
\begin{align*}
\gamma_{b} & =\int dk\,\tilde{\tilde{b}}\left(k\right)^{\dagger}i\partial_{k}\tilde{\tilde{b}}\left(k\right)\\
 & =\int dk\,\tilde{\tilde{b}}\left(k\right)^{\dagger}\left(\mathcal{C}_{2}\theta\right)^{\dagger}\left(\mathcal{C}_{2}\theta\right)i\partial_{k}\tilde{\tilde{b}}\left(k\right)\\
 & =-\int dk\,\tilde{\tilde{b}}\left(k\right)^{\dagger}e^{ik}i\partial_{k}e^{-ik}\tilde{\tilde{b}}\left(k\right)\\
 & =-\gamma_{b}-\int dk\,\partial_{k}k\\
 & =-\gamma_{b}-2\pi,
\end{align*}
from which we see $\gamma_{a}=0$ and $\gamma_{b}=\pi$, and thus we
find that the partial polarization is $\pi$. In particular, this
shows that the nested polarization around $\nu=0,1/2$ will be $\pi$
when there are an odd number of crossings in half the Brillouin zone
on this line.

Let us finally consider two occupied Kramers' pairs with a disconnected
Wilson loop spectrum (see Fig. \ref{fig:wlsbz}(d)-(e)). To calculate
the partial polarization of these bands, let us first consider the
red Kramers pair in isolation. To calculate the partial polarization
we need to again calculate the Berry phase of the red dotted band.
This band does not posses $\mathcal{C}_{2}$ symmetry and hence its
Berry phase will not be quantized. In order to calculate the partial
polarization of the blue bands, we calculate the Berry phase of the
blue dotted band. However, since the blue dotted and the red dotted
bands are related by $\mathcal{C}_{2}$ symmetry, we find for their
Berry phases
\begin{align*}
\gamma_{\mathrm{Red}}^{I} & =-\gamma_{\mathrm{Blue}}^{I},
\end{align*}
and hence the partial polarization, which is the sum of the two, is
zero.

To calculate \emph{nested} partial polarizations, we need to select
symmetric regions centered around $\nu=0$ and $\nu=1/2$. Since there are
two gaps in the spectrum, we have two choices. We can either include
the pair of Kramers pairs, or exclude them from either region. However,
we have just seen that the partial polarization of this set of bands
is zero, and hence either choice will yield the same result. Indeed
for any $N_{F}$, the only choice in selecting a subset of bands centered around
$\nu=0,1/2$, is including or excluding pairs of disconnected bands
such as in Fig. \ref{fig:wlsbz}(e), making the nested partial polarizations
well-defined quantities.

Finally, gapped Wilson loop spectra for arbitrary $N_{F}$ will consist
of linear combinations of the three cases presented here, and since
the partial polarization is an additive quantity, we know how to calculate
it for arbitrary $N_{F}.$

\section{Topological insulator models of the fragile type with $N_F=2$}\label{sec:fragile2}
To show the existence of fragile topological insulating phases in two-dimensional crystals with two-fold rotation and time-reversal symmetries when only one pair of Kramers related pairs are occupied, we start by considering the following two-band model of a Chern insulator with $\mathcal{C}=+2$,

\begin{align*}
\mathcal{H} & =\left\{ -t\left[1+\cos\left(k_{y}\right)+\cos\left(2k_{x}\right)\right]-t_{2}\cos\left(k_{x}\right)\right\} \tau_{x}\\
 & \left\{ -t\left[-\sin\left(k_{y}\right)-\sin\left(2k_{x}\right)\right] -t_{2}\sin\left(k_{x}\right) \right\} \tau_{y}\\
 & -t_{3}\sin\left(2k_{x}\right)\tau_{z}
\end{align*}
where $t,t_{2},t_{3}$ are hopping amplitudes, and $\tau_{i}$ are
Pauli matrices representing an internal degree of freedom. We now
add its time-reversal partner, and couple them with 
\begin{align*}
\mathcal{H}_{R} & =-i\lambda\left\{ \left[-\frac{1}{2}\sin\left(k_{x}\right)+\sin\left(k_{y}\right)\right]i\tau_{x}s_{x}\right.\\
 & \left.+\left[\frac{1}{2}-\frac{1}{2}\cos\left(k_{x}\right)+\cos\left(k_{y}\right)\right]i\tau_{y}s_{x}\right\} ,
\end{align*}
where $\lambda$ is a hopping amplitude and $s_{i}$ are Pauli matrices
acting in spin-space.
This model consists of two time-reversed copies of Chern insulators
with Chern numbers $\mathcal{C}=\pm2$, and $\mathcal{C}_{2}$ symmetry. Taken
together, the Kane-Mele invariant is trivial but the Wilson loop spectra
wind, indicating a fragile topological insulator protected by $\mathcal{C}_{2}\Theta$
symmetry.
The $\mathcal{C}_{2}$ symmetry operator is $\mathcal{C}_{2}=i\tau_{x}\otimes s_{z}$
and the time-reversal operator $\Theta=U\mathcal{K}$, with $U=I_{2}\otimes is_{y}$and
${\mathcal K}$ is complex conjugation. The plots in the main text are for the
parameters $t/t_{2}=0.4$ $t/t_{3}=-1.6$ $\lambda/t=0.15$.

\begin{figure}[tbp]
\includegraphics[width=\columnwidth]{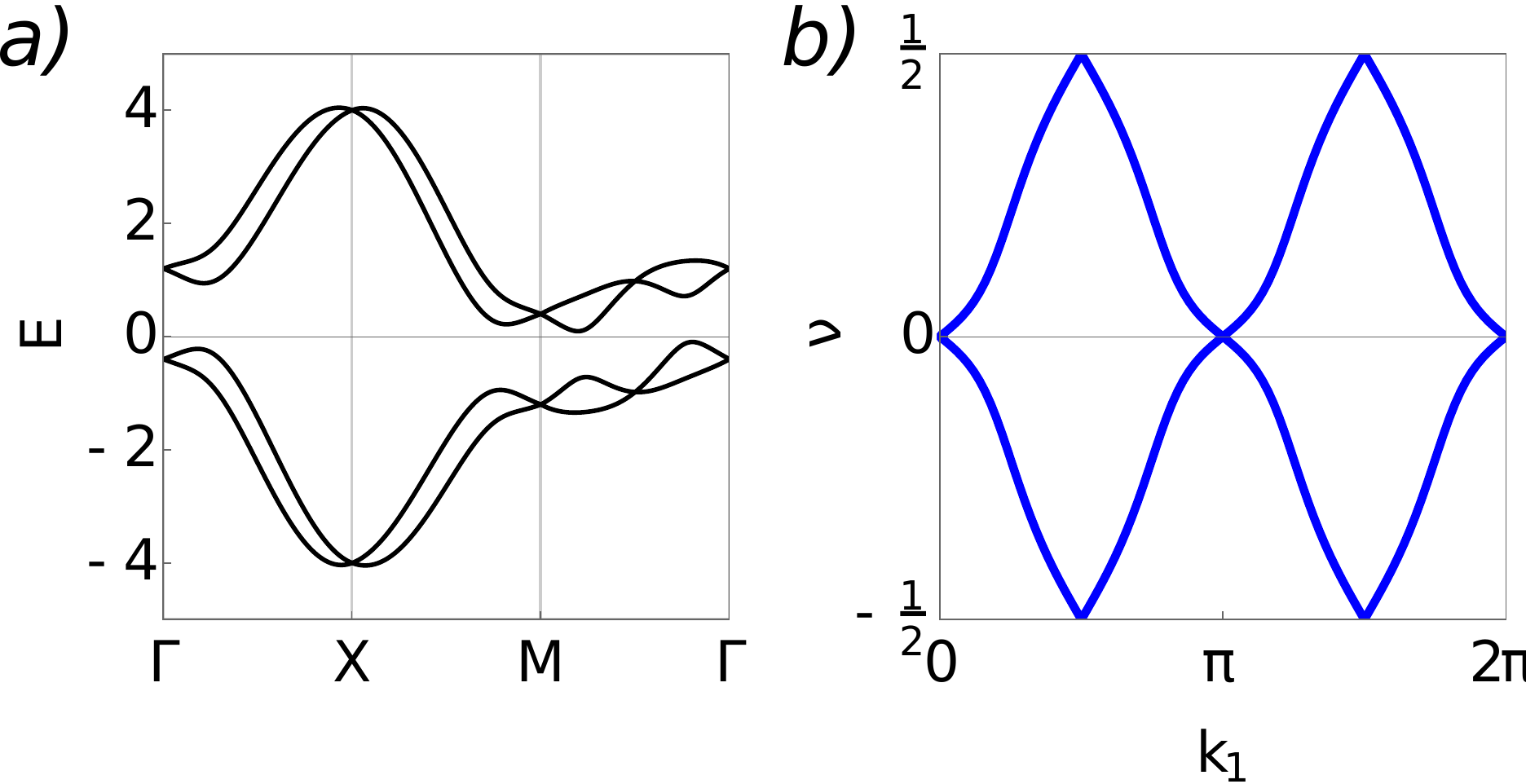}
\caption{Bulk bands (a) and Wilson loop spectrum (b) for the ${\mathcal C}_{4}$ symmetric fragile topological insulator. Plots are for $\epsilon_{1}/t=0.1$,$\epsilon_{2}/t=-0.3$, $t_{2}/t=0.8$, $\lambda/t=0.4$.
 \label{fig:c4}}
\end{figure}

A topological phase of the fragile type can also be obtained in a fourfold rotation symmetric system by considering the following $\mathcal{C}_{4}$ symmetric $\mathcal{C}=+2$ Chern insulator, 

\begin{align*}
\mathcal{H} & =-\epsilon_{1}\left[\cos\left(k_{x}\right)+\cos\left(k_{y}\right)\right]\left(\tau_{0}+\tau_{z}\right)\\
 & -\epsilon_{2}\left[\cos\left(k_{x}\right)+\cos\left(k_{y}\right)\right]\left(\tau_{0}-\tau_{z}\right)\\
 & -2t\left[\cos\left(k_{x}\right)-\cos\left(k_{y}\right)\right]\tau_{x}\\
 & -t_{2}\left[\sin\left(k_{x}\right)\sin\left(k_{y}\right)\right]\tau_{x},
\end{align*}
where $\epsilon_{1},\epsilon_{2},t,t_{2}$ are hopping amplitudes.
We then again add a time-reversal copy and couple them by
\begin{align*}
\mathcal{H}_{\mathrm{mix}} & =-\lambda\left[\sin\left(k_{x}\right)\tau_{0}s_{y}+\sin\left(k_{y}\right)\tau_{0}s_{x}\right].
\end{align*}

The bulk bands and Wilson loop spectrum are plotted in Fig. \ref{fig:c4}.
The $\mathcal{C}_{4}$ operator is represented by $\mathcal{C}_{4}=\tau_{z}\otimes e^{is_{z}/4}$,
and time-reversal by $\Theta=U\mathcal{K}$ with $U=I_{2}\otimes is_{y}$ and $\mathcal{K}$
is complex conjugation.
The symmetry eigenvalues of the occupied bands at $\Gamma$ are $\{e^{i\pi/4},e^{-i\pi/4}\}$, and at the $M$ point $\{-e^{-i\pi/4},-e^{i\pi/4}\}$, which are compatible with a Wannier function centered at the maximal ${\mathcal C}_4$ symmetric position $1b=\left\{1/2, 1/2 \right\}$.

\section{Fragile topological insulators with two occupied Kramers pairs of bands}\label{sec:fragile4}
To construct a fragile topological insulator with two occupied Kramers
pairs, we consider two copies of a quantum spin Hall insulator (the
Kane-Mele model \cite{kan05b}) with broken $\mathcal{C}_{3}$ symmetry on a honeycomb lattice,
\begin{align*}
\mathcal{H}_{\mathrm{KM}}^{\alpha} & =-t^{\alpha}\sum_{\left\langle i,j\right\rangle ,\sigma}c_{i,\sigma}^{\alpha\dagger}c_{j,\sigma}^{\alpha}+\epsilon^{\alpha}\sum_{i,\sigma}c_{i,\sigma}^{\alpha\dagger}c_{j,\sigma}^{\alpha}\\
 & -\left(-1\right)^{\alpha}it_{2}^{\alpha}\sum_{\left\langle \left\langle i,j\right\rangle \right\rangle _{\hat{x}},\sigma}\eta_{ij}c_{i,\sigma}^{\alpha\dagger}c_{j,\sigma}^{\alpha}\\
 & -i\lambda^{\alpha}\sum_{\left\langle i,j\right\rangle ,\sigma,\sigma'}c_{i,\sigma}^{\alpha\dagger}\left(\mathbf{d}\times\mathbf{s}\right)_{z}^{\sigma\sigma'}c_{j,\sigma'}^{\alpha},
\end{align*}
where $\alpha=1,2$ denotes the two copies of the Kane-Mele model,
$t^{\alpha}$ denotes the hopping amplitude, $\left\langle i,j\right\rangle $
the sum over nearest-neighbors, $t_{2}^{\alpha}$ the amplitude of
intrinsic spin-orbit coupling, $\left\langle \left\langle i,j\right\rangle \right\rangle _{\hat{x}}$
the sum over next-nearest neighbors \emph{only }in the $x$-direction,
$\eta_{ij}=+1\left(-1\right)$ for hopping in the clockwise (counter-clockwise)
direction, $\lambda^{\alpha}$ the Rashba amplitude, $\mathbf{d}$
the vector between two sites, $\mathbf{s}$ the vector of Pauli matrices
and $\epsilon^{\alpha}$ an on-site potential. Note that we have only
taken intrinsic spin-orbit coupling along one direction, and hence
we have broken $\mathcal{C}_{3}$ symmetry. 

\begin{figure}
\includegraphics[width=\columnwidth]{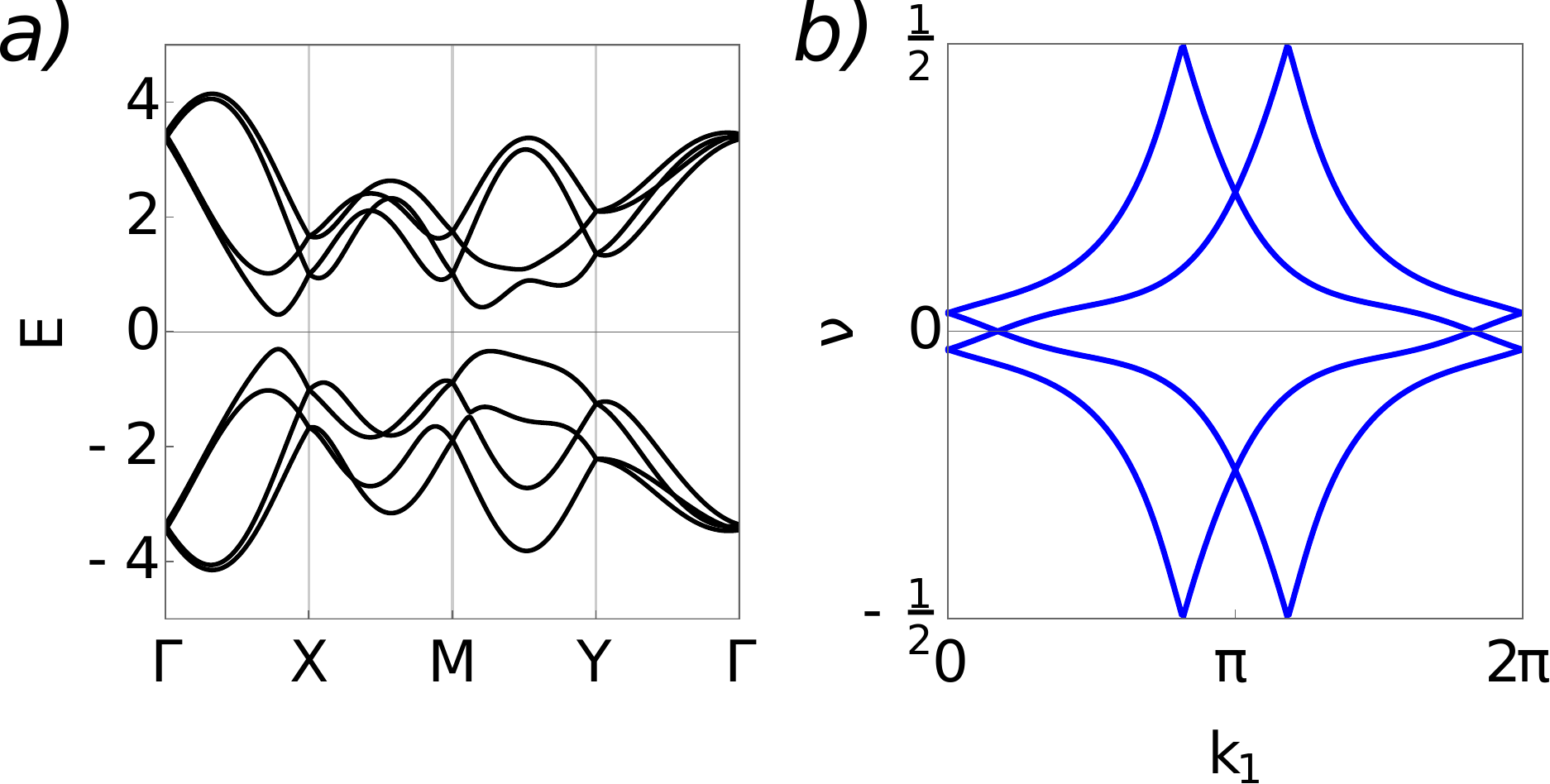}\caption{Bulk band spectrum (a) and Wilson loop spectrum (b) 
of the Hamiltonian
$\mathcal{H}_{p_{x}p_{y}}$ for $\epsilon=0.5$, $t_{\pi}/t_{\sigma}=1.5$,
$t_{2}/t_{\sigma}=3.75$, $\lambda/t_{\sigma}=1.2$ \label{fig:pxpy}}
\end{figure}

We now couple the two copies with the following term
\begin{align*}
\mathcal{H}_{\mathrm{mix}}\left(t_{3}\right) & =-t_{3}\left(\sum_{i,\sigma}c_{i,\sigma}^{2\dagger}c_{i,\sigma}^{1}+\sum_{\left\langle i,j\right\rangle ,\sigma}c_{i,\sigma}^{2\dagger}c_{j,\sigma}^{1}\right.\\
 & \left.+\sum_{\left\langle \left\langle i,j\right\rangle \right\rangle _{\hat{y}},\sigma}c_{i,\sigma}^{2\dagger}c_{j,\sigma}^{1}\right),
\end{align*}
where $\left\langle \left\langle i,j\right\rangle \right\rangle _{\hat{y}}$
denotes next-nearest neighbor hopping along the $\hat{y}$-direction.
When $t_{2}^{1}$ and $t_{2}^{2}$ have a different sign, this Hamiltonian
can be in a fragile topological phase. The time-reversal operators
is $\Theta=U {\mathcal K}$ with $U=i\tau_{0}\sigma_{0}s_{y}$ and the twofold rotation
operator $\mathcal{C}_{2}=i\tau_{0}\sigma_{x}s_{z}$, where $\tau_{i}$, $\sigma_{i}$
and $s_{i}$ are Pauli matrices that act in copy-space, sub-lattice
space and spin-space respectively. The plots in the main text are
made for the parameters $t^{2}/t^{1}=1.1$, $t_{2}^{1}/t^{1}=1.1$,
$t_{2}^{2}/t^{1}=-0.9$, $\epsilon^{1}/t^{1}=-\epsilon^{2}/t^{1}=0.1$,
$\lambda^{1}/t^{1}=\lambda^{2}/t^{1}=0.15$, $t_{3}/t^{1}=0.25.$

A different way to construct a model exhibiting this fragile topological phase is by considering a model of $p_{x, y}$ orbitals on a
honeycomb lattice introduced in Ref. \cite{can18},
\begin{align*}
\mathcal{H}_{p_{x}p_{y}}\left(k\right) & =\left(\begin{array}{cc}
0 & h_{k}\\
h_{k}^{\dagger} & 0
\end{array}\right)+H_{k}^{1},
\end{align*}
with
\begin{align*}
h_{k} & =\frac{1}{2}\left(1+\alpha\,e^{-ik_{2}}+e^{-ik_{1}}\right)\left(t_{\sigma}+t_{\pi}\right)\\
 & -\frac{1}{2}\left(-\frac{1}{2}+\alpha\,e^{-ik_{2}}-\frac{1}{2}e^{-ik_{1}}\right)\left(t_{\sigma}-t_{\pi}\right)\sigma_{z}\\
 & +\frac{\sqrt{3}}{4}\left(-1+e^{-ik_{1}}\right)\left(t_{\sigma}-t_{\pi}\right)\sigma_{x},
\end{align*}
and
\begin{align*}
H_{k}^{1} & =-\frac{t_{2}}{4}\biggl\{\sin\left[i\left(k_{2}-k_{1}\right)\right]+\sin\left[ik_{1}\right]\\
 & \left.-\rho\,\sin\left[ik_{2}\right]\right\} \tau_{z}\otimes\sigma_{y},
\end{align*}
where $t_{\sigma}$ and $t_{\pi}$ are the hopping amplitudes for the
$\sigma$ and $\pi$ pairing, $t_{2}$ is the amplitude of next-nearest-neighbor
hopping and $\sigma_{i}$ and $\tau_{i}$ are Pauli matrices that
act in orbital and sublattice space respectively. $\alpha$ and $\rho$
are two parameters we have introduced to break the $\mathcal{C}_{3}$ symmetry.
For $\alpha=\rho=1$, the $\mathcal{C}_{3}$, symmetry is preserved and our Hamiltonian
is equivalent to the one in Ref.~\cite{can18}. 

We now take two copies of two copies of $\mathcal{H}_{p_{x}p_{y}}\left(k\right)$,
where one copy has spin pointing in the positive $x$-direction, and
the other spin pointing in the negative $x$-direction. In addition,
we shift the momentum along the $x$-direction of the copies in opposite
direction:
\begin{align}
\mathcal{H} & =\mathcal{H}_{p_{x}p_{y}}\left(k-\hat{x}\epsilon\right)\ket{\leftarrow}\bra{\leftarrow}+\mathcal{H}_{p_{x}p_{y}}\left(-k+\hat{x}\ \epsilon\right)^{*}\ket{\rightarrow}\bra{\rightarrow}\nonumber \\
 & +\mathcal{H}_{\mathrm{mix}}\ket{\rightarrow}\bra{\leftarrow},\label{eq:hamspin}
\end{align}
where the spins are mixed by 
\begin{align*}
\mathcal{H}_{\mathrm{mix}} & =-i\lambda\sin\left(k_{x}\right)\tau_{0}\otimes\sigma_{0}.
\end{align*}
This Hamiltonian has $\mathcal{C}_{2}$ and $\Theta$ symmetry, where $\mathcal{C}_{2}=-\tau_{0}\sigma_{x}e^{is_{x}\pi/2}$,
and $\Theta=U\mathcal{K}$ with $U=-i\tau_{0}\sigma_{0}s_{y}$ and $\mathcal{K}$ complex conjugation,
and $\tau_{i},\sigma_{i},s_{i}$ Pauli matrices acting in orbital-space,
sublattice-space and spin-space respectively. Fig. \ref{fig:pxpy}
shows the bulk band spectrum and the Wilson loop spectrum.


\begin{thebibliography}{90}%
\makeatletter
\providecommand \@ifxundefined [1]{%
 \@ifx{#1\undefined}
}%
\providecommand \@ifnum [1]{%
 \ifnum #1\expandafter \@firstoftwo
 \else \expandafter \@secondoftwo
 \fi
}%
\providecommand \@ifx [1]{%
 \ifx #1\expandafter \@firstoftwo
 \else \expandafter \@secondoftwo
 \fi
}%
\providecommand \natexlab [1]{#1}%
\providecommand \enquote  [1]{``#1''}%
\providecommand \bibnamefont  [1]{#1}%
\providecommand \bibfnamefont [1]{#1}%
\providecommand \citenamefont [1]{#1}%
\providecommand \href@noop [0]{\@secondoftwo}%
\providecommand \href [0]{\begingroup \@sanitize@url \@href}%
\providecommand \@href[1]{\@@startlink{#1}\@@href}%
\providecommand \@@href[1]{\endgroup#1\@@endlink}%
\providecommand \@sanitize@url [0]{\catcode `\\12\catcode `\$12\catcode
  `\&12\catcode `\#12\catcode `\^12\catcode `\_12\catcode `\%12\relax}%
\providecommand \@@startlink[1]{}%
\providecommand \@@endlink[0]{}%
\providecommand \url  [0]{\begingroup\@sanitize@url \@url }%
\providecommand \@url [1]{\endgroup\@href {#1}{\urlprefix }}%
\providecommand \urlprefix  [0]{URL }%
\providecommand \Eprint [0]{\href }%
\providecommand \doibase [0]{http://dx.doi.org/}%
\providecommand \selectlanguage [0]{\@gobble}%
\providecommand \bibinfo  [0]{\@secondoftwo}%
\providecommand \bibfield  [0]{\@secondoftwo}%
\providecommand \translation [1]{[#1]}%
\providecommand \BibitemOpen [0]{}%
\providecommand \bibitemStop [0]{}%
\providecommand \bibitemNoStop [0]{.\EOS\space}%
\providecommand \EOS [0]{\spacefactor3000\relax}%
\providecommand \BibitemShut  [1]{\csname bibitem#1\endcsname}%
\let\auto@bib@innerbib\@empty
%</preamble>
\bibitem [{\citenamefont {Klitzing}\ \emph {et~al.}(1980)\citenamefont
  {Klitzing}, \citenamefont {Dorda},\ and\ \citenamefont {Pepper}}]{kli80}%
  \BibitemOpen
  \bibfield  {author} {\bibinfo {author} {\bibfnamefont {K.~v.}\ \bibnamefont
  {Klitzing}}, \bibinfo {author} {\bibfnamefont {G.}~\bibnamefont {Dorda}}, \
  and\ \bibinfo {author} {\bibfnamefont {M.}~\bibnamefont {Pepper}},\
  }\href@noop {} {\bibfield  {journal} {\bibinfo  {journal} {Phys.\ Rev.\
  Lett.}\ }\textbf {\bibinfo {volume} {45}},\ \bibinfo {pages} {494} (\bibinfo
  {year} {1980})}\BibitemShut {NoStop}%
\bibitem [{\citenamefont {Thouless}\ \emph {et~al.}(1982)\citenamefont
  {Thouless}, \citenamefont {Kohmoto}, \citenamefont {Nightingale},\ and\
  \citenamefont {den Nijs}}]{tho82}%
  \BibitemOpen
  \bibfield  {author} {\bibinfo {author} {\bibfnamefont {D.~J.}\ \bibnamefont
  {Thouless}}, \bibinfo {author} {\bibfnamefont {M.}~\bibnamefont {Kohmoto}},
  \bibinfo {author} {\bibfnamefont {M.~P.}\ \bibnamefont {Nightingale}}, \ and\
  \bibinfo {author} {\bibfnamefont {M.}~\bibnamefont {den Nijs}},\ }\href@noop
  {} {\bibfield  {journal} {\bibinfo  {journal} {Phys.\ Rev.\ Lett.}\ }\textbf
  {\bibinfo {volume} {49}},\ \bibinfo {pages} {405} (\bibinfo {year}
  {1982})}\BibitemShut {NoStop}%
\bibitem [{\citenamefont {Halperin}(1982)}]{hal82}%
  \BibitemOpen
  \bibfield  {author} {\bibinfo {author} {\bibfnamefont {B.~I.}\ \bibnamefont
  {Halperin}},\ }\href@noop {} {\bibfield  {journal} {\bibinfo  {journal}
  {Phys.\ Rev.\ B}\ }\textbf {\bibinfo {volume} {25}},\ \bibinfo {pages} {2185}
  (\bibinfo {year} {1982})}\BibitemShut {NoStop}%
\bibitem [{\citenamefont {Kohmoto}(1985)}]{koh85}%
  \BibitemOpen
  \bibfield  {author} {\bibinfo {author} {\bibfnamefont {M.}~\bibnamefont
  {Kohmoto}},\ }\href@noop {} {\bibfield  {journal} {\bibinfo  {journal} {Ann.
  Phys. (N.Y.)}\ }\textbf {\bibinfo {volume} {160}},\ \bibinfo {pages} {343}
  (\bibinfo {year} {1985})}\BibitemShut {NoStop}%
\bibitem [{\citenamefont {Kane}\ and\ \citenamefont
  {Mele}(2005{\natexlab{a}})}]{kan05b}%
  \BibitemOpen
  \bibfield  {author} {\bibinfo {author} {\bibfnamefont {C.~L.}\ \bibnamefont
  {Kane}}\ and\ \bibinfo {author} {\bibfnamefont {E.~J.}\ \bibnamefont
  {Mele}},\ }\href {\doibase 10.1103/PhysRevLett.95.226801} {\bibfield
  {journal} {\bibinfo  {journal} {Phys. Rev. Lett.}\ }\textbf {\bibinfo
  {volume} {95}},\ \bibinfo {pages} {226801} (\bibinfo {year}
  {2005}{\natexlab{a}})}\BibitemShut {NoStop}%
\bibitem [{\citenamefont {Kane}\ and\ \citenamefont
  {Mele}(2005{\natexlab{b}})}]{kan05}%
  \BibitemOpen
  \bibfield  {author} {\bibinfo {author} {\bibfnamefont {C.~L.}\ \bibnamefont
  {Kane}}\ and\ \bibinfo {author} {\bibfnamefont {E.~J.}\ \bibnamefont
  {Mele}},\ }\href {\doibase 10.1103/PhysRevLett.95.146802} {\bibfield
  {journal} {\bibinfo  {journal} {Phys. Rev. Lett.}\ }\textbf {\bibinfo
  {volume} {95}},\ \bibinfo {pages} {146802} (\bibinfo {year}
  {2005}{\natexlab{b}})}\BibitemShut {NoStop}%
\bibitem [{\citenamefont {Bernevig}\ \emph {et~al.}(2006)\citenamefont
  {Bernevig}, \citenamefont {Hughes},\ and\ \citenamefont {Zhang}}]{ber06}%
  \BibitemOpen
  \bibfield  {author} {\bibinfo {author} {\bibfnamefont {B.~A.}\ \bibnamefont
  {Bernevig}}, \bibinfo {author} {\bibfnamefont {T.~L.}\ \bibnamefont
  {Hughes}}, \ and\ \bibinfo {author} {\bibfnamefont {S.-C.}\ \bibnamefont
  {Zhang}},\ }\href@noop {} {\bibfield  {journal} {\bibinfo  {journal}
  {Science}\ }\textbf {\bibinfo {volume} {314}},\ \bibinfo {pages} {1757}
  (\bibinfo {year} {2006})}\BibitemShut {NoStop}%
\bibitem [{\citenamefont {K{\"o}nig}\ \emph {et~al.}(2007)\citenamefont
  {K{\"o}nig}, \citenamefont {Wiedmann}, \citenamefont {Br{\"u}ne},
  \citenamefont {Roth}, \citenamefont {Buhmann}, \citenamefont {Molenkamp},
  \citenamefont {Qi},\ and\ \citenamefont {Zhang}}]{mol07}%
  \BibitemOpen
  \bibfield  {author} {\bibinfo {author} {\bibfnamefont {M.}~\bibnamefont
  {K{\"o}nig}}, \bibinfo {author} {\bibfnamefont {S.}~\bibnamefont {Wiedmann}},
  \bibinfo {author} {\bibfnamefont {C.}~\bibnamefont {Br{\"u}ne}}, \bibinfo
  {author} {\bibfnamefont {A.}~\bibnamefont {Roth}}, \bibinfo {author}
  {\bibfnamefont {H.}~\bibnamefont {Buhmann}}, \bibinfo {author} {\bibfnamefont
  {L.~W.}\ \bibnamefont {Molenkamp}}, \bibinfo {author} {\bibfnamefont {X.-L.}\
  \bibnamefont {Qi}}, \ and\ \bibinfo {author} {\bibfnamefont {S.-C.}\
  \bibnamefont {Zhang}},\ }\href@noop {} {\bibfield  {journal} {\bibinfo
  {journal} {Science}\ }\textbf {\bibinfo {volume} {318}},\ \bibinfo {pages}
  {766} (\bibinfo {year} {2007})}\BibitemShut {NoStop}%
\bibitem [{\citenamefont {Zhang}\ \emph {et~al.}(2009)\citenamefont {Zhang},
  \citenamefont {Liu}, \citenamefont {Qi}, \citenamefont {Dai}, \citenamefont
  {Fang},\ and\ \citenamefont {Zhang}}]{zha09}%
  \BibitemOpen
  \bibfield  {author} {\bibinfo {author} {\bibfnamefont {H.}~\bibnamefont
  {Zhang}}, \bibinfo {author} {\bibfnamefont {C.-X.}\ \bibnamefont {Liu}},
  \bibinfo {author} {\bibfnamefont {X.-L.}\ \bibnamefont {Qi}}, \bibinfo
  {author} {\bibfnamefont {X.}~\bibnamefont {Dai}}, \bibinfo {author}
  {\bibfnamefont {Z.}~\bibnamefont {Fang}}, \ and\ \bibinfo {author}
  {\bibfnamefont {S.-C.}\ \bibnamefont {Zhang}},\ }\href@noop {} {\bibfield
  {journal} {\bibinfo  {journal} {Nat. Phys.}\ }\textbf {\bibinfo {volume}
  {5}},\ \bibinfo {pages} {438} (\bibinfo {year} {2009})}\BibitemShut {NoStop}%
\bibitem [{\citenamefont {Fu}\ \emph {et~al.}(2007)\citenamefont {Fu},
  \citenamefont {Kane},\ and\ \citenamefont {Mele}}]{fu07}%
  \BibitemOpen
  \bibfield  {author} {\bibinfo {author} {\bibfnamefont {L.}~\bibnamefont
  {Fu}}, \bibinfo {author} {\bibfnamefont {C.~L.}\ \bibnamefont {Kane}}, \ and\
  \bibinfo {author} {\bibfnamefont {E.~J.}\ \bibnamefont {Mele}},\ }\href@noop
  {} {\bibfield  {journal} {\bibinfo  {journal} {Phys.\ Rev.\ Lett.}\ }\textbf
  {\bibinfo {volume} {98}},\ \bibinfo {pages} {106803} (\bibinfo {year}
  {2007})}\BibitemShut {NoStop}%
\bibitem [{\citenamefont {Br{\"u}ne}\ \emph {et~al.}(2011)\citenamefont
  {Br{\"u}ne}, \citenamefont {Liu}, \citenamefont {Novik}, \citenamefont
  {Hankiewicz}, \citenamefont {Buhmann}, \citenamefont {Chen}, \citenamefont
  {Qi}, \citenamefont {Shen}, \citenamefont {Zhang},\ and\ \citenamefont
  {Molenkamp}}]{bru11}%
  \BibitemOpen
  \bibfield  {author} {\bibinfo {author} {\bibfnamefont {C.}~\bibnamefont
  {Br{\"u}ne}}, \bibinfo {author} {\bibfnamefont {C.}~\bibnamefont {Liu}},
  \bibinfo {author} {\bibfnamefont {E.}~\bibnamefont {Novik}}, \bibinfo
  {author} {\bibfnamefont {E.}~\bibnamefont {Hankiewicz}}, \bibinfo {author}
  {\bibfnamefont {H.}~\bibnamefont {Buhmann}}, \bibinfo {author} {\bibfnamefont
  {Y.}~\bibnamefont {Chen}}, \bibinfo {author} {\bibfnamefont {X.}~\bibnamefont
  {Qi}}, \bibinfo {author} {\bibfnamefont {Z.}~\bibnamefont {Shen}}, \bibinfo
  {author} {\bibfnamefont {S.}~\bibnamefont {Zhang}}, \ and\ \bibinfo {author}
  {\bibfnamefont {L.}~\bibnamefont {Molenkamp}},\ }\href@noop {} {\bibfield
  {journal} {\bibinfo  {journal} {Phys.\ Rev.\ Lett.}\ }\textbf {\bibinfo
  {volume} {106}},\ \bibinfo {pages} {126803} (\bibinfo {year}
  {2011})}\BibitemShut {NoStop}%
\bibitem [{\citenamefont {Rasche}\ \emph {et~al.}(2013)\citenamefont {Rasche},
  \citenamefont {Isaeva}, \citenamefont {Ruck}, \citenamefont {Borisenko},
  \citenamefont {Zabolotnyy}, \citenamefont {B{\"u}chner}, \citenamefont
  {Koepernik}, \citenamefont {Ortix}, \citenamefont {Richter},\ and\
  \citenamefont {Van Den~Brink}}]{ras13}%
  \BibitemOpen
  \bibfield  {author} {\bibinfo {author} {\bibfnamefont {B.}~\bibnamefont
  {Rasche}}, \bibinfo {author} {\bibfnamefont {A.}~\bibnamefont {Isaeva}},
  \bibinfo {author} {\bibfnamefont {M.}~\bibnamefont {Ruck}}, \bibinfo {author}
  {\bibfnamefont {S.}~\bibnamefont {Borisenko}}, \bibinfo {author}
  {\bibfnamefont {V.}~\bibnamefont {Zabolotnyy}}, \bibinfo {author}
  {\bibfnamefont {B.}~\bibnamefont {B{\"u}chner}}, \bibinfo {author}
  {\bibfnamefont {K.}~\bibnamefont {Koepernik}}, \bibinfo {author}
  {\bibfnamefont {C.}~\bibnamefont {Ortix}}, \bibinfo {author} {\bibfnamefont
  {M.}~\bibnamefont {Richter}}, \ and\ \bibinfo {author} {\bibfnamefont
  {J.}~\bibnamefont {Van Den~Brink}},\ }\href@noop {} {\bibfield  {journal}
  {\bibinfo  {journal} {Nat. Mat.}\ }\textbf {\bibinfo {volume} {12}},\
  \bibinfo {pages} {422} (\bibinfo {year} {2013})}\BibitemShut {NoStop}%
\bibitem [{\citenamefont {Mourik}\ \emph {et~al.}(2012)\citenamefont {Mourik},
  \citenamefont {Zuo}, \citenamefont {Frolov}, \citenamefont {Plissard},
  \citenamefont {Bakkers},\ and\ \citenamefont {Kouwenhoven}}]{mou12}%
  \BibitemOpen
  \bibfield  {author} {\bibinfo {author} {\bibfnamefont {V.}~\bibnamefont
  {Mourik}}, \bibinfo {author} {\bibfnamefont {K.}~\bibnamefont {Zuo}},
  \bibinfo {author} {\bibfnamefont {S.~M.}\ \bibnamefont {Frolov}}, \bibinfo
  {author} {\bibfnamefont {S.}~\bibnamefont {Plissard}}, \bibinfo {author}
  {\bibfnamefont {E.~P.}\ \bibnamefont {Bakkers}}, \ and\ \bibinfo {author}
  {\bibfnamefont {L.~P.}\ \bibnamefont {Kouwenhoven}},\ }\href@noop {}
  {\bibfield  {journal} {\bibinfo  {journal} {Science}\ }\textbf {\bibinfo
  {volume} {336}},\ \bibinfo {pages} {1003} (\bibinfo {year}
  {2012})}\BibitemShut {NoStop}%
\bibitem [{\citenamefont {Lutchyn}\ \emph {et~al.}(2010)\citenamefont
  {Lutchyn}, \citenamefont {Sau},\ and\ \citenamefont {Sarma}}]{lut10}%
  \BibitemOpen
  \bibfield  {author} {\bibinfo {author} {\bibfnamefont {R.~M.}\ \bibnamefont
  {Lutchyn}}, \bibinfo {author} {\bibfnamefont {J.~D.}\ \bibnamefont {Sau}}, \
  and\ \bibinfo {author} {\bibfnamefont {S.~D.}\ \bibnamefont {Sarma}},\
  }\href@noop {} {\bibfield  {journal} {\bibinfo  {journal} {Phys.\ Rev.\
  Lett.}\ }\textbf {\bibinfo {volume} {105}},\ \bibinfo {pages} {077001}
  (\bibinfo {year} {2010})}\BibitemShut {NoStop}%
\bibitem [{\citenamefont {Fu}\ and\ \citenamefont {Kane}(2008)}]{fu08}%
  \BibitemOpen
  \bibfield  {author} {\bibinfo {author} {\bibfnamefont {L.}~\bibnamefont
  {Fu}}\ and\ \bibinfo {author} {\bibfnamefont {C.~L.}\ \bibnamefont {Kane}},\
  }\href@noop {} {\bibfield  {journal} {\bibinfo  {journal} {Phys.\ Rev.\
  Lett.}\ }\textbf {\bibinfo {volume} {100}},\ \bibinfo {pages} {096407}
  (\bibinfo {year} {2008})}\BibitemShut {NoStop}%
\bibitem [{\citenamefont {Beenakker}(2013)}]{bee13}%
  \BibitemOpen
  \bibfield  {author} {\bibinfo {author} {\bibfnamefont {C.}~\bibnamefont
  {Beenakker}},\ }\href@noop {} {\bibfield  {journal} {\bibinfo  {journal}
  {Ann. Rev. Cond. Mat. Phys.}\ }\textbf {\bibinfo {volume} {4}},\ \bibinfo
  {pages} {113} (\bibinfo {year} {2013})}\BibitemShut {NoStop}%
\bibitem [{\citenamefont {Alicea}(2012)}]{ali12}%
  \BibitemOpen
  \bibfield  {author} {\bibinfo {author} {\bibfnamefont {J.}~\bibnamefont
  {Alicea}},\ }\href@noop {} {\bibfield  {journal} {\bibinfo  {journal} {Rep.
  Prog. in Phys.}\ }\textbf {\bibinfo {volume} {75}},\ \bibinfo {pages}
  {076501} (\bibinfo {year} {2012})}\BibitemShut {NoStop}%
\bibitem [{\citenamefont {Armitage}\ \emph {et~al.}(2018)\citenamefont
  {Armitage}, \citenamefont {Mele},\ and\ \citenamefont {Vishwanath}}]{arm18}%
  \BibitemOpen
  \bibfield  {author} {\bibinfo {author} {\bibfnamefont {N.}~\bibnamefont
  {Armitage}}, \bibinfo {author} {\bibfnamefont {E.}~\bibnamefont {Mele}}, \
  and\ \bibinfo {author} {\bibfnamefont {A.}~\bibnamefont {Vishwanath}},\
  }\href@noop {} {\bibfield  {journal} {\bibinfo  {journal} {Rev.\ Mod.\
  Phys.}\ }\textbf {\bibinfo {volume} {90}},\ \bibinfo {pages} {015001}
  (\bibinfo {year} {2018})}\BibitemShut {NoStop}%
\bibitem [{\citenamefont {Huang}\ \emph {et~al.}(2015)\citenamefont {Huang},
  \citenamefont {Xu}, \citenamefont {Belopolski}, \citenamefont {Lee},
  \citenamefont {Chang}, \citenamefont {Wang}, \citenamefont {Alidoust},
  \citenamefont {Bian}, \citenamefont {Neupane}, \citenamefont {Zhang} \emph
  {et~al.}}]{hua15}%
  \BibitemOpen
  \bibfield  {author} {\bibinfo {author} {\bibfnamefont {S.-M.}\ \bibnamefont
  {Huang}}, \bibinfo {author} {\bibfnamefont {S.-Y.}\ \bibnamefont {Xu}},
  \bibinfo {author} {\bibfnamefont {I.}~\bibnamefont {Belopolski}}, \bibinfo
  {author} {\bibfnamefont {C.-C.}\ \bibnamefont {Lee}}, \bibinfo {author}
  {\bibfnamefont {G.}~\bibnamefont {Chang}}, \bibinfo {author} {\bibfnamefont
  {B.}~\bibnamefont {Wang}}, \bibinfo {author} {\bibfnamefont {N.}~\bibnamefont
  {Alidoust}}, \bibinfo {author} {\bibfnamefont {G.}~\bibnamefont {Bian}},
  \bibinfo {author} {\bibfnamefont {M.}~\bibnamefont {Neupane}}, \bibinfo
  {author} {\bibfnamefont {C.}~\bibnamefont {Zhang}},  \emph {et~al.},\
  }\href@noop {} {\bibfield  {journal} {\bibinfo  {journal} {Nat. Comm.}\
  }\textbf {\bibinfo {volume} {6}},\ \bibinfo {pages} {7373} (\bibinfo {year}
  {2015})}\BibitemShut {NoStop}%
\bibitem [{\citenamefont {Lv}\ \emph {et~al.}(2015)\citenamefont {Lv},
  \citenamefont {Weng}, \citenamefont {Fu}, \citenamefont {Wang}, \citenamefont
  {Miao}, \citenamefont {Ma}, \citenamefont {Richard}, \citenamefont {Huang},
  \citenamefont {Zhao}, \citenamefont {Chen} \emph {et~al.}}]{lv15}%
  \BibitemOpen
  \bibfield  {author} {\bibinfo {author} {\bibfnamefont {B.}~\bibnamefont
  {Lv}}, \bibinfo {author} {\bibfnamefont {H.}~\bibnamefont {Weng}}, \bibinfo
  {author} {\bibfnamefont {B.}~\bibnamefont {Fu}}, \bibinfo {author}
  {\bibfnamefont {X.}~\bibnamefont {Wang}}, \bibinfo {author} {\bibfnamefont
  {H.}~\bibnamefont {Miao}}, \bibinfo {author} {\bibfnamefont {J.}~\bibnamefont
  {Ma}}, \bibinfo {author} {\bibfnamefont {P.}~\bibnamefont {Richard}},
  \bibinfo {author} {\bibfnamefont {X.}~\bibnamefont {Huang}}, \bibinfo
  {author} {\bibfnamefont {L.}~\bibnamefont {Zhao}}, \bibinfo {author}
  {\bibfnamefont {G.}~\bibnamefont {Chen}},  \emph {et~al.},\ }\href@noop {}
  {\bibfield  {journal} {\bibinfo  {journal} {Phys.\ Rev. \ X}\ }\textbf
  {\bibinfo {volume} {5}},\ \bibinfo {pages} {031013} (\bibinfo {year}
  {2015})}\BibitemShut {NoStop}%
\bibitem [{\citenamefont {Xu}\ \emph {et~al.}(2015)\citenamefont {Xu},
  \citenamefont {Belopolski}, \citenamefont {Alidoust}, \citenamefont
  {Neupane}, \citenamefont {Bian}, \citenamefont {Zhang}, \citenamefont
  {Sankar}, \citenamefont {Chang}, \citenamefont {Yuan}, \citenamefont {Lee}
  \emph {et~al.}}]{xu15}%
  \BibitemOpen
  \bibfield  {author} {\bibinfo {author} {\bibfnamefont {S.-Y.}\ \bibnamefont
  {Xu}}, \bibinfo {author} {\bibfnamefont {I.}~\bibnamefont {Belopolski}},
  \bibinfo {author} {\bibfnamefont {N.}~\bibnamefont {Alidoust}}, \bibinfo
  {author} {\bibfnamefont {M.}~\bibnamefont {Neupane}}, \bibinfo {author}
  {\bibfnamefont {G.}~\bibnamefont {Bian}}, \bibinfo {author} {\bibfnamefont
  {C.}~\bibnamefont {Zhang}}, \bibinfo {author} {\bibfnamefont
  {R.}~\bibnamefont {Sankar}}, \bibinfo {author} {\bibfnamefont
  {G.}~\bibnamefont {Chang}}, \bibinfo {author} {\bibfnamefont
  {Z.}~\bibnamefont {Yuan}}, \bibinfo {author} {\bibfnamefont {C.-C.}\
  \bibnamefont {Lee}},  \emph {et~al.},\ }\href@noop {} {\bibfield  {journal}
  {\bibinfo  {journal} {Science}\ }\textbf {\bibinfo {volume} {349}},\ \bibinfo
  {pages} {613} (\bibinfo {year} {2015})}\BibitemShut {NoStop}%
\bibitem [{\citenamefont {Haubold}\ \emph {et~al.}(2017)\citenamefont
  {Haubold}, \citenamefont {Koepernik}, \citenamefont {Efremov}, \citenamefont
  {Khim}, \citenamefont {Fedorov}, \citenamefont {Kushnirenko}, \citenamefont
  {van~den Brink}, \citenamefont {Wurmehl}, \citenamefont {B{\"u}chner},
  \citenamefont {Kim} \emph {et~al.}}]{hau17}%
  \BibitemOpen
  \bibfield  {author} {\bibinfo {author} {\bibfnamefont {E.}~\bibnamefont
  {Haubold}}, \bibinfo {author} {\bibfnamefont {K.}~\bibnamefont {Koepernik}},
  \bibinfo {author} {\bibfnamefont {D.}~\bibnamefont {Efremov}}, \bibinfo
  {author} {\bibfnamefont {S.}~\bibnamefont {Khim}}, \bibinfo {author}
  {\bibfnamefont {A.}~\bibnamefont {Fedorov}}, \bibinfo {author} {\bibfnamefont
  {Y.}~\bibnamefont {Kushnirenko}}, \bibinfo {author} {\bibfnamefont
  {J.}~\bibnamefont {van~den Brink}}, \bibinfo {author} {\bibfnamefont
  {S.}~\bibnamefont {Wurmehl}}, \bibinfo {author} {\bibfnamefont
  {B.}~\bibnamefont {B{\"u}chner}}, \bibinfo {author} {\bibfnamefont
  {T.}~\bibnamefont {Kim}},  \emph {et~al.},\ }\href@noop {} {\bibfield
  {journal} {\bibinfo  {journal} {Phys.\ Rev.\ B}\ }\textbf {\bibinfo {volume}
  {95}},\ \bibinfo {pages} {241108} (\bibinfo {year} {2017})}\BibitemShut
  {NoStop}%
\bibitem [{\citenamefont {Lau}\ \emph {et~al.}(2017)\citenamefont {Lau},
  \citenamefont {Koepernik}, \citenamefont {van~den Brink},\ and\ \citenamefont
  {Ortix}}]{lau17}%
  \BibitemOpen
  \bibfield  {author} {\bibinfo {author} {\bibfnamefont {A.}~\bibnamefont
  {Lau}}, \bibinfo {author} {\bibfnamefont {K.}~\bibnamefont {Koepernik}},
  \bibinfo {author} {\bibfnamefont {J.}~\bibnamefont {van~den Brink}}, \ and\
  \bibinfo {author} {\bibfnamefont {C.}~\bibnamefont {Ortix}},\ }\href
  {\doibase 10.1103/PhysRevLett.119.076801} {\bibfield  {journal} {\bibinfo
  {journal} {Phys. Rev. Lett.}\ }\textbf {\bibinfo {volume} {119}},\ \bibinfo
  {pages} {076801} (\bibinfo {year} {2017})}\BibitemShut {NoStop}%
\bibitem [{\citenamefont {Wan}\ \emph {et~al.}(2011)\citenamefont {Wan},
  \citenamefont {Turner}, \citenamefont {Vishwanath},\ and\ \citenamefont
  {Savrasov}}]{wan11}%
  \BibitemOpen
  \bibfield  {author} {\bibinfo {author} {\bibfnamefont {X.}~\bibnamefont
  {Wan}}, \bibinfo {author} {\bibfnamefont {A.~M.}\ \bibnamefont {Turner}},
  \bibinfo {author} {\bibfnamefont {A.}~\bibnamefont {Vishwanath}}, \ and\
  \bibinfo {author} {\bibfnamefont {S.~Y.}\ \bibnamefont {Savrasov}},\ }\href
  {\doibase 10.1103/PhysRevB.83.205101} {\bibfield  {journal} {\bibinfo
  {journal} {Phys. Rev. B}\ }\textbf {\bibinfo {volume} {83}},\ \bibinfo
  {pages} {205101} (\bibinfo {year} {2011})}\BibitemShut {NoStop}%
\bibitem [{\citenamefont {Burkov}\ and\ \citenamefont {Balents}(2011)}]{bur11}%
  \BibitemOpen
  \bibfield  {author} {\bibinfo {author} {\bibfnamefont {A.~A.}\ \bibnamefont
  {Burkov}}\ and\ \bibinfo {author} {\bibfnamefont {L.}~\bibnamefont
  {Balents}},\ }\href {\doibase 10.1103/PhysRevLett.107.127205} {\bibfield
  {journal} {\bibinfo  {journal} {Phys. Rev. Lett.}\ }\textbf {\bibinfo
  {volume} {107}},\ \bibinfo {pages} {127205} (\bibinfo {year}
  {2011})}\BibitemShut {NoStop}%
\bibitem [{\citenamefont {Zyuzin}\ \emph {et~al.}(2012)\citenamefont {Zyuzin},
  \citenamefont {Wu},\ and\ \citenamefont {Burkov}}]{zyu12}%
  \BibitemOpen
  \bibfield  {author} {\bibinfo {author} {\bibfnamefont {A.~A.}\ \bibnamefont
  {Zyuzin}}, \bibinfo {author} {\bibfnamefont {S.}~\bibnamefont {Wu}}, \ and\
  \bibinfo {author} {\bibfnamefont {A.~A.}\ \bibnamefont {Burkov}},\ }\href
  {\doibase 10.1103/PhysRevB.85.165110} {\bibfield  {journal} {\bibinfo
  {journal} {Phys. Rev. B}\ }\textbf {\bibinfo {volume} {85}},\ \bibinfo
  {pages} {165110} (\bibinfo {year} {2012})}\BibitemShut {NoStop}%
\bibitem [{\citenamefont {Lau}\ and\ \citenamefont {Ortix}(2019)}]{lau19}%
  \BibitemOpen
  \bibfield  {author} {\bibinfo {author} {\bibfnamefont {A.}~\bibnamefont
  {Lau}}\ and\ \bibinfo {author} {\bibfnamefont {C.}~\bibnamefont {Ortix}},\
  }\href {\doibase 10.1103/PhysRevLett.122.186801} {\bibfield  {journal}
  {\bibinfo  {journal} {Phys. Rev. Lett.}\ }\textbf {\bibinfo {volume} {122}},\
  \bibinfo {pages} {186801} (\bibinfo {year} {2019})}\BibitemShut {NoStop}%
\bibitem [{\citenamefont {Ojanen}(2013)}]{oja13}%
  \BibitemOpen
  \bibfield  {author} {\bibinfo {author} {\bibfnamefont {T.}~\bibnamefont
  {Ojanen}},\ }\href {\doibase 10.1103/PhysRevB.87.245112} {\bibfield
  {journal} {\bibinfo  {journal} {Phys. Rev. B}\ }\textbf {\bibinfo {volume}
  {87}},\ \bibinfo {pages} {245112} (\bibinfo {year} {2013})}\BibitemShut
  {NoStop}%
\bibitem [{\citenamefont {Soluyanov}\ \emph {et~al.}(2015)\citenamefont
  {Soluyanov}, \citenamefont {Gresch}, \citenamefont {Wang}, \citenamefont
  {Wu}, \citenamefont {Troyer}, \citenamefont {Dai},\ and\ \citenamefont
  {Bernevig}}]{sol15}%
  \BibitemOpen
  \bibfield  {author} {\bibinfo {author} {\bibfnamefont {A.~A.}\ \bibnamefont
  {Soluyanov}}, \bibinfo {author} {\bibfnamefont {D.}~\bibnamefont {Gresch}},
  \bibinfo {author} {\bibfnamefont {Z.}~\bibnamefont {Wang}}, \bibinfo {author}
  {\bibfnamefont {Q.}~\bibnamefont {Wu}}, \bibinfo {author} {\bibfnamefont
  {M.}~\bibnamefont {Troyer}}, \bibinfo {author} {\bibfnamefont
  {X.}~\bibnamefont {Dai}}, \ and\ \bibinfo {author} {\bibfnamefont {B.~A.}\
  \bibnamefont {Bernevig}},\ }\href@noop {} {\bibfield  {journal} {\bibinfo
  {journal} {Nature}\ }\textbf {\bibinfo {volume} {527}},\ \bibinfo {pages}
  {495} (\bibinfo {year} {2015})}\BibitemShut {NoStop}%
\bibitem [{\citenamefont {Altland}\ and\ \citenamefont
  {Zirnbauer}(1997)}]{alt97}%
  \BibitemOpen
  \bibfield  {author} {\bibinfo {author} {\bibfnamefont {A.}~\bibnamefont
  {Altland}}\ and\ \bibinfo {author} {\bibfnamefont {M.~R.}\ \bibnamefont
  {Zirnbauer}},\ }\href@noop {} {\bibfield  {journal} {\bibinfo  {journal}
  {Phys.\ Rev.\ B}\ }\textbf {\bibinfo {volume} {55}},\ \bibinfo {pages} {1142}
  (\bibinfo {year} {1997})}\BibitemShut {NoStop}%
\bibitem [{\citenamefont {Schnyder}\ \emph {et~al.}(2008)\citenamefont
  {Schnyder}, \citenamefont {Ryu}, \citenamefont {Furusaki},\ and\
  \citenamefont {Ludwig}}]{sch08}%
  \BibitemOpen
  \bibfield  {author} {\bibinfo {author} {\bibfnamefont {A.~P.}\ \bibnamefont
  {Schnyder}}, \bibinfo {author} {\bibfnamefont {S.}~\bibnamefont {Ryu}},
  \bibinfo {author} {\bibfnamefont {A.}~\bibnamefont {Furusaki}}, \ and\
  \bibinfo {author} {\bibfnamefont {A.~W.}\ \bibnamefont {Ludwig}},\
  }\href@noop {} {\bibfield  {journal} {\bibinfo  {journal} {Phys.\ Rev.\ B}\
  }\textbf {\bibinfo {volume} {78}},\ \bibinfo {pages} {195125} (\bibinfo
  {year} {2008})}\BibitemShut {NoStop}%
\bibitem [{\citenamefont {Kitaev}(2009)}]{kit09}%
  \BibitemOpen
  \bibfield  {author} {\bibinfo {author} {\bibfnamefont {A.}~\bibnamefont
  {Kitaev}},\ }in\ \href@noop {} {\emph {\bibinfo {booktitle} {AIP Conference
  Proceedings}}},\ Vol.\ \bibinfo {volume} {1134}\ (\bibinfo {organization}
  {AIP},\ \bibinfo {year} {2009})\ pp.\ \bibinfo {pages} {22--30}\BibitemShut
  {NoStop}%
\bibitem [{\citenamefont {Hasan}\ and\ \citenamefont {Kane}(2010)}]{has10}%
  \BibitemOpen
  \bibfield  {author} {\bibinfo {author} {\bibfnamefont {M.~Z.}\ \bibnamefont
  {Hasan}}\ and\ \bibinfo {author} {\bibfnamefont {C.~L.}\ \bibnamefont
  {Kane}},\ }\href@noop {} {\bibfield  {journal} {\bibinfo  {journal} {Rev.\
  Mod.\ Phys.}\ }\textbf {\bibinfo {volume} {82}},\ \bibinfo {pages} {3045}
  (\bibinfo {year} {2010})}\BibitemShut {NoStop}%
\bibitem [{\citenamefont {Qi}\ and\ \citenamefont {Zhang}(2011)}]{qi11}%
  \BibitemOpen
  \bibfield  {author} {\bibinfo {author} {\bibfnamefont {X.-L.}\ \bibnamefont
  {Qi}}\ and\ \bibinfo {author} {\bibfnamefont {S.-C.}\ \bibnamefont {Zhang}},\
  }\href@noop {} {\bibfield  {journal} {\bibinfo  {journal} {Rev.\ Mod.\
  Phys.}\ }\textbf {\bibinfo {volume} {83}},\ \bibinfo {pages} {1057} (\bibinfo
  {year} {2011})}\BibitemShut {NoStop}%
\bibitem [{\citenamefont {Fu}(2011)}]{fu11}%
  \BibitemOpen
  \bibfield  {author} {\bibinfo {author} {\bibfnamefont {L.}~\bibnamefont
  {Fu}},\ }\href@noop {} {\bibfield  {journal} {\bibinfo  {journal} {Phys.\
  Rev.\ Lett.}\ }\textbf {\bibinfo {volume} {106}},\ \bibinfo {pages} {106802}
  (\bibinfo {year} {2011})}\BibitemShut {NoStop}%
\bibitem [{\citenamefont {Hsieh}\ \emph {et~al.}(2012)\citenamefont {Hsieh},
  \citenamefont {Lin}, \citenamefont {Liu}, \citenamefont {Duan}, \citenamefont
  {Bansil},\ and\ \citenamefont {Fu}}]{hsi12}%
  \BibitemOpen
  \bibfield  {author} {\bibinfo {author} {\bibfnamefont {T.~H.}\ \bibnamefont
  {Hsieh}}, \bibinfo {author} {\bibfnamefont {H.}~\bibnamefont {Lin}}, \bibinfo
  {author} {\bibfnamefont {J.}~\bibnamefont {Liu}}, \bibinfo {author}
  {\bibfnamefont {W.}~\bibnamefont {Duan}}, \bibinfo {author} {\bibfnamefont
  {A.}~\bibnamefont {Bansil}}, \ and\ \bibinfo {author} {\bibfnamefont
  {L.}~\bibnamefont {Fu}},\ }\href@noop {} {\bibfield  {journal} {\bibinfo
  {journal} {Nat. Comm.}\ }\textbf {\bibinfo {volume} {3}},\ \bibinfo {pages}
  {982} (\bibinfo {year} {2012})}\BibitemShut {NoStop}%
\bibitem [{\citenamefont {Liu}\ \emph {et~al.}(2014)\citenamefont {Liu},
  \citenamefont {Hsieh}, \citenamefont {Wei}, \citenamefont {Duan},
  \citenamefont {Moodera},\ and\ \citenamefont {Fu}}]{liu14b}%
  \BibitemOpen
  \bibfield  {author} {\bibinfo {author} {\bibfnamefont {J.}~\bibnamefont
  {Liu}}, \bibinfo {author} {\bibfnamefont {T.~H.}\ \bibnamefont {Hsieh}},
  \bibinfo {author} {\bibfnamefont {P.}~\bibnamefont {Wei}}, \bibinfo {author}
  {\bibfnamefont {W.}~\bibnamefont {Duan}}, \bibinfo {author} {\bibfnamefont
  {J.}~\bibnamefont {Moodera}}, \ and\ \bibinfo {author} {\bibfnamefont
  {L.}~\bibnamefont {Fu}},\ }\href@noop {} {\bibfield  {journal} {\bibinfo
  {journal} {Nat. Mat.}\ }\textbf {\bibinfo {volume} {13}},\ \bibinfo {pages}
  {178} (\bibinfo {year} {2014})}\BibitemShut {NoStop}%
\bibitem [{\citenamefont {Hsieh}\ \emph {et~al.}(2014)\citenamefont {Hsieh},
  \citenamefont {Liu},\ and\ \citenamefont {Fu}}]{hsi14}%
  \BibitemOpen
  \bibfield  {author} {\bibinfo {author} {\bibfnamefont {T.~H.}\ \bibnamefont
  {Hsieh}}, \bibinfo {author} {\bibfnamefont {J.}~\bibnamefont {Liu}}, \ and\
  \bibinfo {author} {\bibfnamefont {L.}~\bibnamefont {Fu}},\ }\href@noop {}
  {\bibfield  {journal} {\bibinfo  {journal} {Phys.\ Rev.\ B}\ }\textbf
  {\bibinfo {volume} {90}},\ \bibinfo {pages} {081112} (\bibinfo {year}
  {2014})}\BibitemShut {NoStop}%
\bibitem [{\citenamefont {Fang}\ and\ \citenamefont {Fu}(2017)}]{fan17}%
  \BibitemOpen
  \bibfield  {author} {\bibinfo {author} {\bibfnamefont {C.}~\bibnamefont
  {Fang}}\ and\ \bibinfo {author} {\bibfnamefont {L.}~\bibnamefont {Fu}},\
  }\href@noop {} {\bibfield  {journal} {\bibinfo  {journal} {arXiv preprint
  arXiv:1709.01929}\ } (\bibinfo {year} {2017})}\BibitemShut {NoStop}%
\bibitem [{\citenamefont {Khalaf}\ \emph {et~al.}(2018)\citenamefont {Khalaf},
  \citenamefont {Po}, \citenamefont {Vishwanath},\ and\ \citenamefont
  {Watanabe}}]{kha18}%
  \BibitemOpen
  \bibfield  {author} {\bibinfo {author} {\bibfnamefont {E.}~\bibnamefont
  {Khalaf}}, \bibinfo {author} {\bibfnamefont {H.~C.}\ \bibnamefont {Po}},
  \bibinfo {author} {\bibfnamefont {A.}~\bibnamefont {Vishwanath}}, \ and\
  \bibinfo {author} {\bibfnamefont {H.}~\bibnamefont {Watanabe}},\ }\href@noop
  {} {\bibfield  {journal} {\bibinfo  {journal} {Phys.\ Rev. \ X}\ }\textbf
  {\bibinfo {volume} {8}},\ \bibinfo {pages} {031070} (\bibinfo {year}
  {2018})}\BibitemShut {NoStop}%
\bibitem [{\citenamefont {Schindler}\ \emph
  {et~al.}(2018{\natexlab{a}})\citenamefont {Schindler}, \citenamefont {Cook},
  \citenamefont {Vergniory}, \citenamefont {Wang}, \citenamefont {Parkin},
  \citenamefont {Bernevig},\ and\ \citenamefont {Neupert}}]{sch18}%
  \BibitemOpen
  \bibfield  {author} {\bibinfo {author} {\bibfnamefont {F.}~\bibnamefont
  {Schindler}}, \bibinfo {author} {\bibfnamefont {A.~M.}\ \bibnamefont {Cook}},
  \bibinfo {author} {\bibfnamefont {M.~G.}\ \bibnamefont {Vergniory}}, \bibinfo
  {author} {\bibfnamefont {Z.}~\bibnamefont {Wang}}, \bibinfo {author}
  {\bibfnamefont {S.~S.}\ \bibnamefont {Parkin}}, \bibinfo {author}
  {\bibfnamefont {B.~A.}\ \bibnamefont {Bernevig}}, \ and\ \bibinfo {author}
  {\bibfnamefont {T.}~\bibnamefont {Neupert}},\ }\href@noop {} {\bibfield
  {journal} {\bibinfo  {journal} {Sci. Adv.}\ }\textbf {\bibinfo {volume}
  {4}},\ \bibinfo {pages} {eaat0346} (\bibinfo {year}
  {2018}{\natexlab{a}})}\BibitemShut {NoStop}%
\bibitem [{\citenamefont {Benalcazar}\ \emph
  {et~al.}(2017{\natexlab{a}})\citenamefont {Benalcazar}, \citenamefont
  {Bernevig},\ and\ \citenamefont {Hughes}}]{ben17}%
  \BibitemOpen
  \bibfield  {author} {\bibinfo {author} {\bibfnamefont {W.~A.}\ \bibnamefont
  {Benalcazar}}, \bibinfo {author} {\bibfnamefont {B.~A.}\ \bibnamefont
  {Bernevig}}, \ and\ \bibinfo {author} {\bibfnamefont {T.~L.}\ \bibnamefont
  {Hughes}},\ }\href@noop {} {\bibfield  {journal} {\bibinfo  {journal}
  {Science}\ }\textbf {\bibinfo {volume} {357}},\ \bibinfo {pages} {61}
  (\bibinfo {year} {2017}{\natexlab{a}})}\BibitemShut {NoStop}%
\bibitem [{\citenamefont {Benalcazar}\ \emph
  {et~al.}(2017{\natexlab{b}})\citenamefont {Benalcazar}, \citenamefont
  {Bernevig},\ and\ \citenamefont {Hughes}}]{ben17-2}%
  \BibitemOpen
  \bibfield  {author} {\bibinfo {author} {\bibfnamefont {W.~A.}\ \bibnamefont
  {Benalcazar}}, \bibinfo {author} {\bibfnamefont {B.~A.}\ \bibnamefont
  {Bernevig}}, \ and\ \bibinfo {author} {\bibfnamefont {T.~L.}\ \bibnamefont
  {Hughes}},\ }\href@noop {} {\bibfield  {journal} {\bibinfo  {journal}
  {Physical Review B}\ }\textbf {\bibinfo {volume} {96}},\ \bibinfo {pages}
  {245115} (\bibinfo {year} {2017}{\natexlab{b}})}\BibitemShut {NoStop}%
\bibitem [{\citenamefont {Schindler}\ \emph
  {et~al.}(2018{\natexlab{b}})\citenamefont {Schindler}, \citenamefont {Wang},
  \citenamefont {Vergniory}, \citenamefont {Cook}, \citenamefont {Murani},
  \citenamefont {Sengupta}, \citenamefont {Kasumov}, \citenamefont {Deblock},
  \citenamefont {Jeon}, \citenamefont {Drozdov} \emph {et~al.}}]{sch18b}%
  \BibitemOpen
  \bibfield  {author} {\bibinfo {author} {\bibfnamefont {F.}~\bibnamefont
  {Schindler}}, \bibinfo {author} {\bibfnamefont {Z.}~\bibnamefont {Wang}},
  \bibinfo {author} {\bibfnamefont {M.~G.}\ \bibnamefont {Vergniory}}, \bibinfo
  {author} {\bibfnamefont {A.~M.}\ \bibnamefont {Cook}}, \bibinfo {author}
  {\bibfnamefont {A.}~\bibnamefont {Murani}}, \bibinfo {author} {\bibfnamefont
  {S.}~\bibnamefont {Sengupta}}, \bibinfo {author} {\bibfnamefont {A.~Y.}\
  \bibnamefont {Kasumov}}, \bibinfo {author} {\bibfnamefont {R.}~\bibnamefont
  {Deblock}}, \bibinfo {author} {\bibfnamefont {S.}~\bibnamefont {Jeon}},
  \bibinfo {author} {\bibfnamefont {I.}~\bibnamefont {Drozdov}},  \emph
  {et~al.},\ }\href@noop {} {\bibfield  {journal} {\bibinfo  {journal} {Nat.
  Phys.}\ }\textbf {\bibinfo {volume} {14}},\ \bibinfo {pages} {918} (\bibinfo
  {year} {2018}{\natexlab{b}})}\BibitemShut {NoStop}%
\bibitem [{\citenamefont {Geier}\ \emph {et~al.}(2018)\citenamefont {Geier},
  \citenamefont {Trifunovic}, \citenamefont {Hoskam},\ and\ \citenamefont
  {Brouwer}}]{gei18}%
  \BibitemOpen
  \bibfield  {author} {\bibinfo {author} {\bibfnamefont {M.}~\bibnamefont
  {Geier}}, \bibinfo {author} {\bibfnamefont {L.}~\bibnamefont {Trifunovic}},
  \bibinfo {author} {\bibfnamefont {M.}~\bibnamefont {Hoskam}}, \ and\ \bibinfo
  {author} {\bibfnamefont {P.~W.}\ \bibnamefont {Brouwer}},\ }\href {\doibase
  10.1103/PhysRevB.97.205135} {\bibfield  {journal} {\bibinfo  {journal} {Phys.
  Rev. B}\ }\textbf {\bibinfo {volume} {97}},\ \bibinfo {pages} {205135}
  (\bibinfo {year} {2018})}\BibitemShut {NoStop}%
\bibitem [{\citenamefont {Peterson}\ \emph {et~al.}(2018)\citenamefont
  {Peterson}, \citenamefont {Benalcazar}, \citenamefont {Hughes},\ and\
  \citenamefont {Bahl}}]{pet18}%
  \BibitemOpen
  \bibfield  {author} {\bibinfo {author} {\bibfnamefont {C.~W.}\ \bibnamefont
  {Peterson}}, \bibinfo {author} {\bibfnamefont {W.~A.}\ \bibnamefont
  {Benalcazar}}, \bibinfo {author} {\bibfnamefont {T.~L.}\ \bibnamefont
  {Hughes}}, \ and\ \bibinfo {author} {\bibfnamefont {G.}~\bibnamefont
  {Bahl}},\ }\href@noop {} {\bibfield  {journal} {\bibinfo  {journal} {Nature}\
  }\textbf {\bibinfo {volume} {555}},\ \bibinfo {pages} {346} (\bibinfo {year}
  {2018})}\BibitemShut {NoStop}%
\bibitem [{\citenamefont {Serra-Garcia}\ \emph {et~al.}(2018)\citenamefont
  {Serra-Garcia}, \citenamefont {Peri}, \citenamefont {S{\"u}sstrunk},
  \citenamefont {Bilal}, \citenamefont {Larsen}, \citenamefont {Villanueva},\
  and\ \citenamefont {Huber}}]{ser18}%
  \BibitemOpen
  \bibfield  {author} {\bibinfo {author} {\bibfnamefont {M.}~\bibnamefont
  {Serra-Garcia}}, \bibinfo {author} {\bibfnamefont {V.}~\bibnamefont {Peri}},
  \bibinfo {author} {\bibfnamefont {R.}~\bibnamefont {S{\"u}sstrunk}}, \bibinfo
  {author} {\bibfnamefont {O.~R.}\ \bibnamefont {Bilal}}, \bibinfo {author}
  {\bibfnamefont {T.}~\bibnamefont {Larsen}}, \bibinfo {author} {\bibfnamefont
  {L.~G.}\ \bibnamefont {Villanueva}}, \ and\ \bibinfo {author} {\bibfnamefont
  {S.~D.}\ \bibnamefont {Huber}},\ }\href@noop {} {\bibfield  {journal}
  {\bibinfo  {journal} {Nature}\ }\textbf {\bibinfo {volume} {555}},\ \bibinfo
  {pages} {342} (\bibinfo {year} {2018})}\BibitemShut {NoStop}%
\bibitem [{\citenamefont {Khalaf}(2018)}]{kha18b}%
  \BibitemOpen
  \bibfield  {author} {\bibinfo {author} {\bibfnamefont {E.}~\bibnamefont
  {Khalaf}},\ }\href {\doibase 10.1103/PhysRevB.97.205136} {\bibfield
  {journal} {\bibinfo  {journal} {Phys. Rev. B}\ }\textbf {\bibinfo {volume}
  {97}},\ \bibinfo {pages} {205136} (\bibinfo {year} {2018})}\BibitemShut
  {NoStop}%
\bibitem [{\citenamefont {Ezawa}(2018{\natexlab{a}})}]{eza18}%
  \BibitemOpen
  \bibfield  {author} {\bibinfo {author} {\bibfnamefont {M.}~\bibnamefont
  {Ezawa}},\ }\href {\doibase 10.1103/PhysRevLett.120.026801} {\bibfield
  {journal} {\bibinfo  {journal} {Phys. Rev. Lett.}\ }\textbf {\bibinfo
  {volume} {120}},\ \bibinfo {pages} {026801} (\bibinfo {year}
  {2018}{\natexlab{a}})}\BibitemShut {NoStop}%
\bibitem [{\citenamefont {Ezawa}(2018{\natexlab{b}})}]{eza18b}%
  \BibitemOpen
  \bibfield  {author} {\bibinfo {author} {\bibfnamefont {M.}~\bibnamefont
  {Ezawa}},\ }\href {\doibase 10.1103/PhysRevB.97.155305} {\bibfield  {journal}
  {\bibinfo  {journal} {Phys. Rev. B}\ }\textbf {\bibinfo {volume} {97}},\
  \bibinfo {pages} {155305} (\bibinfo {year} {2018}{\natexlab{b}})}\BibitemShut
  {NoStop}%
\bibitem [{\citenamefont {Langbehn}\ \emph {et~al.}(2017)\citenamefont
  {Langbehn}, \citenamefont {Peng}, \citenamefont {Trifunovic}, \citenamefont
  {von Oppen},\ and\ \citenamefont {Brouwer}}]{lan17}%
  \BibitemOpen
  \bibfield  {author} {\bibinfo {author} {\bibfnamefont {J.}~\bibnamefont
  {Langbehn}}, \bibinfo {author} {\bibfnamefont {Y.}~\bibnamefont {Peng}},
  \bibinfo {author} {\bibfnamefont {L.}~\bibnamefont {Trifunovic}}, \bibinfo
  {author} {\bibfnamefont {F.}~\bibnamefont {von Oppen}}, \ and\ \bibinfo
  {author} {\bibfnamefont {P.~W.}\ \bibnamefont {Brouwer}},\ }\href {\doibase
  10.1103/PhysRevLett.119.246401} {\bibfield  {journal} {\bibinfo  {journal}
  {Phys. Rev. Lett.}\ }\textbf {\bibinfo {volume} {119}},\ \bibinfo {pages}
  {246401} (\bibinfo {year} {2017})}\BibitemShut {NoStop}%
\bibitem [{\citenamefont {Sitte}\ \emph {et~al.}(2012)\citenamefont {Sitte},
  \citenamefont {Rosch}, \citenamefont {Altman},\ and\ \citenamefont
  {Fritz}}]{sit12}%
  \BibitemOpen
  \bibfield  {author} {\bibinfo {author} {\bibfnamefont {M.}~\bibnamefont
  {Sitte}}, \bibinfo {author} {\bibfnamefont {A.}~\bibnamefont {Rosch}},
  \bibinfo {author} {\bibfnamefont {E.}~\bibnamefont {Altman}}, \ and\ \bibinfo
  {author} {\bibfnamefont {L.}~\bibnamefont {Fritz}},\ }\href {\doibase
  10.1103/PhysRevLett.108.126807} {\bibfield  {journal} {\bibinfo  {journal}
  {Phys. Rev. Lett.}\ }\textbf {\bibinfo {volume} {108}},\ \bibinfo {pages}
  {126807} (\bibinfo {year} {2012})}\BibitemShut {NoStop}%
\bibitem [{\citenamefont {Song}\ \emph {et~al.}(2017)\citenamefont {Song},
  \citenamefont {Fang},\ and\ \citenamefont {Fang}}]{son17}%
  \BibitemOpen
  \bibfield  {author} {\bibinfo {author} {\bibfnamefont {Z.}~\bibnamefont
  {Song}}, \bibinfo {author} {\bibfnamefont {Z.}~\bibnamefont {Fang}}, \ and\
  \bibinfo {author} {\bibfnamefont {C.}~\bibnamefont {Fang}},\ }\href@noop {}
  {\bibfield  {journal} {\bibinfo  {journal} {Phys. Rev. Lett.}\ }\textbf
  {\bibinfo {volume} {119}},\ \bibinfo {pages} {246402} (\bibinfo {year}
  {2017})}\BibitemShut {NoStop}%
\bibitem [{\citenamefont {Imhof}\ \emph {et~al.}(2018)\citenamefont {Imhof},
  \citenamefont {Berger}, \citenamefont {Bayer}, \citenamefont {Brehm},
  \citenamefont {Molenkamp}, \citenamefont {Kiessling}, \citenamefont
  {Schindler}, \citenamefont {Lee}, \citenamefont {Greiter}, \citenamefont
  {Neupert} \emph {et~al.}}]{imh18}%
  \BibitemOpen
  \bibfield  {author} {\bibinfo {author} {\bibfnamefont {S.}~\bibnamefont
  {Imhof}}, \bibinfo {author} {\bibfnamefont {C.}~\bibnamefont {Berger}},
  \bibinfo {author} {\bibfnamefont {F.}~\bibnamefont {Bayer}}, \bibinfo
  {author} {\bibfnamefont {J.}~\bibnamefont {Brehm}}, \bibinfo {author}
  {\bibfnamefont {L.~W.}\ \bibnamefont {Molenkamp}}, \bibinfo {author}
  {\bibfnamefont {T.}~\bibnamefont {Kiessling}}, \bibinfo {author}
  {\bibfnamefont {F.}~\bibnamefont {Schindler}}, \bibinfo {author}
  {\bibfnamefont {C.~H.}\ \bibnamefont {Lee}}, \bibinfo {author} {\bibfnamefont
  {M.}~\bibnamefont {Greiter}}, \bibinfo {author} {\bibfnamefont
  {T.}~\bibnamefont {Neupert}},  \emph {et~al.},\ }\href@noop {} {\bibfield
  {journal} {\bibinfo  {journal} {Nat. Phys.}\ }\textbf {\bibinfo {volume}
  {14}},\ \bibinfo {pages} {925} (\bibinfo {year} {2018})}\BibitemShut
  {NoStop}%
\bibitem [{\citenamefont {Xu}\ \emph {et~al.}(2017)\citenamefont {Xu},
  \citenamefont {Xue},\ and\ \citenamefont {Wan}}]{xu17}%
  \BibitemOpen
  \bibfield  {author} {\bibinfo {author} {\bibfnamefont {Y.}~\bibnamefont
  {Xu}}, \bibinfo {author} {\bibfnamefont {R.}~\bibnamefont {Xue}}, \ and\
  \bibinfo {author} {\bibfnamefont {S.}~\bibnamefont {Wan}},\ }\href@noop {}
  {\bibfield  {journal} {\bibinfo  {journal} {arXiv preprint arXiv:1711.09202}\
  } (\bibinfo {year} {2017})}\BibitemShut {NoStop}%
\bibitem [{\citenamefont {Hsu}\ \emph {et~al.}(2018)\citenamefont {Hsu},
  \citenamefont {Stano}, \citenamefont {Klinovaja},\ and\ \citenamefont
  {Loss}}]{hsu18}%
  \BibitemOpen
  \bibfield  {author} {\bibinfo {author} {\bibfnamefont {C.-H.}\ \bibnamefont
  {Hsu}}, \bibinfo {author} {\bibfnamefont {P.}~\bibnamefont {Stano}}, \bibinfo
  {author} {\bibfnamefont {J.}~\bibnamefont {Klinovaja}}, \ and\ \bibinfo
  {author} {\bibfnamefont {D.}~\bibnamefont {Loss}},\ }\href@noop {} {\bibfield
   {journal} {\bibinfo  {journal} {Phys.\ Rev.\ Lett.}\ }\textbf {\bibinfo
  {volume} {121}},\ \bibinfo {pages} {196801} (\bibinfo {year}
  {2018})}\BibitemShut {NoStop}%
\bibitem [{\citenamefont {van Miert}\ and\ \citenamefont
  {Ortix}(2018{\natexlab{a}})}]{mie18}%
  \BibitemOpen
  \bibfield  {author} {\bibinfo {author} {\bibfnamefont {G.}~\bibnamefont {van
  Miert}}\ and\ \bibinfo {author} {\bibfnamefont {C.}~\bibnamefont {Ortix}},\
  }\href@noop {} {\bibfield  {journal} {\bibinfo  {journal} {Phys. Rev. B}\
  }\textbf {\bibinfo {volume} {98}},\ \bibinfo {pages} {081110} (\bibinfo
  {year} {2018}{\natexlab{a}})}\BibitemShut {NoStop}%
\bibitem [{\citenamefont {Kooi}\ \emph {et~al.}(2018)\citenamefont {Kooi},
  \citenamefont {Van~Miert},\ and\ \citenamefont {Ortix}}]{koo18}%
  \BibitemOpen
  \bibfield  {author} {\bibinfo {author} {\bibfnamefont {S.~H.}\ \bibnamefont
  {Kooi}}, \bibinfo {author} {\bibfnamefont {G.}~\bibnamefont {Van~Miert}}, \
  and\ \bibinfo {author} {\bibfnamefont {C.}~\bibnamefont {Ortix}},\
  }\href@noop {} {\bibfield  {journal} {\bibinfo  {journal} {Phys.\ Rev.\ B}\
  }\textbf {\bibinfo {volume} {98}},\ \bibinfo {pages} {245102} (\bibinfo
  {year} {2018})}\BibitemShut {NoStop}%
\bibitem [{\citenamefont {Soluyanov}\ and\ \citenamefont
  {Vanderbilt}(2011)}]{sol11}%
  \BibitemOpen
  \bibfield  {author} {\bibinfo {author} {\bibfnamefont {A.~A.}\ \bibnamefont
  {Soluyanov}}\ and\ \bibinfo {author} {\bibfnamefont {D.}~\bibnamefont
  {Vanderbilt}},\ }\href@noop {} {\bibfield  {journal} {\bibinfo  {journal}
  {Phys.\ Rev.\ B}\ }\textbf {\bibinfo {volume} {83}},\ \bibinfo {pages}
  {035108} (\bibinfo {year} {2011})}\BibitemShut {NoStop}%
\bibitem [{\citenamefont {Yu}\ \emph {et~al.}(2011)\citenamefont {Yu},
  \citenamefont {Qi}, \citenamefont {Bernevig}, \citenamefont {Fang},\ and\
  \citenamefont {Dai}}]{yu11}%
  \BibitemOpen
  \bibfield  {author} {\bibinfo {author} {\bibfnamefont {R.}~\bibnamefont
  {Yu}}, \bibinfo {author} {\bibfnamefont {X.~L.}\ \bibnamefont {Qi}}, \bibinfo
  {author} {\bibfnamefont {A.}~\bibnamefont {Bernevig}}, \bibinfo {author}
  {\bibfnamefont {Z.}~\bibnamefont {Fang}}, \ and\ \bibinfo {author}
  {\bibfnamefont {X.}~\bibnamefont {Dai}},\ }\href {\doibase
  10.1103/PhysRevB.84.075119} {\bibfield  {journal} {\bibinfo  {journal} {Phys.
  Rev. B}\ }\textbf {\bibinfo {volume} {84}},\ \bibinfo {pages} {075119}
  (\bibinfo {year} {2011})}\BibitemShut {NoStop}%
\bibitem [{\citenamefont {Bradlyn}\ \emph {et~al.}(2017)\citenamefont
  {Bradlyn}, \citenamefont {Elcoro}, \citenamefont {Cano}, \citenamefont
  {Vergniory}, \citenamefont {Wang}, \citenamefont {Felser}, \citenamefont
  {Aroyo},\ and\ \citenamefont {Bernevig}}]{bra17}%
  \BibitemOpen
  \bibfield  {author} {\bibinfo {author} {\bibfnamefont {B.}~\bibnamefont
  {Bradlyn}}, \bibinfo {author} {\bibfnamefont {L.}~\bibnamefont {Elcoro}},
  \bibinfo {author} {\bibfnamefont {J.}~\bibnamefont {Cano}}, \bibinfo {author}
  {\bibfnamefont {M.~G.}\ \bibnamefont {Vergniory}}, \bibinfo {author}
  {\bibfnamefont {Z.}~\bibnamefont {Wang}}, \bibinfo {author} {\bibfnamefont
  {C.}~\bibnamefont {Felser}}, \bibinfo {author} {\bibfnamefont {M.~I.}\
  \bibnamefont {Aroyo}}, \ and\ \bibinfo {author} {\bibfnamefont {B.~A.}\
  \bibnamefont {Bernevig}},\ }\href@noop {} {\bibfield  {journal} {\bibinfo
  {journal} {Nature}\ }\textbf {\bibinfo {volume} {547}},\ \bibinfo {pages}
  {298} (\bibinfo {year} {2017})}\BibitemShut {NoStop}%
\bibitem [{\citenamefont {Po}\ \emph {et~al.}(2017)\citenamefont {Po},
  \citenamefont {Vishwanath},\ and\ \citenamefont {Watanabe}}]{po17}%
  \BibitemOpen
  \bibfield  {author} {\bibinfo {author} {\bibfnamefont {H.~C.}\ \bibnamefont
  {Po}}, \bibinfo {author} {\bibfnamefont {A.}~\bibnamefont {Vishwanath}}, \
  and\ \bibinfo {author} {\bibfnamefont {H.}~\bibnamefont {Watanabe}},\
  }\href@noop {} {\bibfield  {journal} {\bibinfo  {journal} {Nature
  Communications}\ }\textbf {\bibinfo {volume} {8}},\ \bibinfo {pages} {50}
  (\bibinfo {year} {2017})}\BibitemShut {NoStop}%
\bibitem [{\citenamefont {Kruthoff}\ \emph {et~al.}(2017)\citenamefont
  {Kruthoff}, \citenamefont {de~Boer}, \citenamefont {van Wezel}, \citenamefont
  {Kane},\ and\ \citenamefont {Slager}}]{kru17b}%
  \BibitemOpen
  \bibfield  {author} {\bibinfo {author} {\bibfnamefont {J.}~\bibnamefont
  {Kruthoff}}, \bibinfo {author} {\bibfnamefont {J.}~\bibnamefont {de~Boer}},
  \bibinfo {author} {\bibfnamefont {J.}~\bibnamefont {van Wezel}}, \bibinfo
  {author} {\bibfnamefont {C.~L.}\ \bibnamefont {Kane}}, \ and\ \bibinfo
  {author} {\bibfnamefont {R.-J.}\ \bibnamefont {Slager}},\ }\href@noop {}
  {\bibfield  {journal} {\bibinfo  {journal} {Phys.\ Rev. \ X}\ }\textbf
  {\bibinfo {volume} {7}},\ \bibinfo {pages} {041069} (\bibinfo {year}
  {2017})}\BibitemShut {NoStop}%
\bibitem [{\citenamefont {Song}\ \emph {et~al.}(2018)\citenamefont {Song},
  \citenamefont {Zhang}, \citenamefont {Fang},\ and\ \citenamefont
  {Fang}}]{son18}%
  \BibitemOpen
  \bibfield  {author} {\bibinfo {author} {\bibfnamefont {Z.}~\bibnamefont
  {Song}}, \bibinfo {author} {\bibfnamefont {T.}~\bibnamefont {Zhang}},
  \bibinfo {author} {\bibfnamefont {Z.}~\bibnamefont {Fang}}, \ and\ \bibinfo
  {author} {\bibfnamefont {C.}~\bibnamefont {Fang}},\ }\href@noop {} {\bibfield
   {journal} {\bibinfo  {journal} {Nat. Comm.}\ }\textbf {\bibinfo {volume}
  {9}},\ \bibinfo {pages} {3530} (\bibinfo {year} {2018})}\BibitemShut
  {NoStop}%
\bibitem [{\citenamefont {Zhang}\ \emph {et~al.}(2019)\citenamefont {Zhang},
  \citenamefont {Jiang}, \citenamefont {Song}, \citenamefont {Huang},
  \citenamefont {He}, \citenamefont {Fang}, \citenamefont {Weng},\ and\
  \citenamefont {Fang}}]{zha19}%
  \BibitemOpen
  \bibfield  {author} {\bibinfo {author} {\bibfnamefont {T.}~\bibnamefont
  {Zhang}}, \bibinfo {author} {\bibfnamefont {Y.}~\bibnamefont {Jiang}},
  \bibinfo {author} {\bibfnamefont {Z.}~\bibnamefont {Song}}, \bibinfo {author}
  {\bibfnamefont {H.}~\bibnamefont {Huang}}, \bibinfo {author} {\bibfnamefont
  {Y.}~\bibnamefont {He}}, \bibinfo {author} {\bibfnamefont {Z.}~\bibnamefont
  {Fang}}, \bibinfo {author} {\bibfnamefont {H.}~\bibnamefont {Weng}}, \ and\
  \bibinfo {author} {\bibfnamefont {C.}~\bibnamefont {Fang}},\ }\href@noop {}
  {\bibfield  {journal} {\bibinfo  {journal} {Nature}\ }\textbf {\bibinfo
  {volume} {566}},\ \bibinfo {pages} {475} (\bibinfo {year}
  {2019})}\BibitemShut {NoStop}%
\bibitem [{\citenamefont {Vergniory}\ \emph {et~al.}(2019)\citenamefont
  {Vergniory}, \citenamefont {Elcoro}, \citenamefont {Felser}, \citenamefont
  {Regnault}, \citenamefont {Bernevig},\ and\ \citenamefont {Wang}}]{ver19}%
  \BibitemOpen
  \bibfield  {author} {\bibinfo {author} {\bibfnamefont {M.}~\bibnamefont
  {Vergniory}}, \bibinfo {author} {\bibfnamefont {L.}~\bibnamefont {Elcoro}},
  \bibinfo {author} {\bibfnamefont {C.}~\bibnamefont {Felser}}, \bibinfo
  {author} {\bibfnamefont {N.}~\bibnamefont {Regnault}}, \bibinfo {author}
  {\bibfnamefont {B.~A.}\ \bibnamefont {Bernevig}}, \ and\ \bibinfo {author}
  {\bibfnamefont {Z.}~\bibnamefont {Wang}},\ }\href@noop {} {\bibfield
  {journal} {\bibinfo  {journal} {Nature}\ }\textbf {\bibinfo {volume} {566}},\
  \bibinfo {pages} {480} (\bibinfo {year} {2019})}\BibitemShut {NoStop}%
\bibitem [{\citenamefont {Tang}\ \emph {et~al.}(2019)\citenamefont {Tang},
  \citenamefont {Po}, \citenamefont {Vishwanath},\ and\ \citenamefont
  {Wan}}]{tan19}%
  \BibitemOpen
  \bibfield  {author} {\bibinfo {author} {\bibfnamefont {F.}~\bibnamefont
  {Tang}}, \bibinfo {author} {\bibfnamefont {H.~C.}\ \bibnamefont {Po}},
  \bibinfo {author} {\bibfnamefont {A.}~\bibnamefont {Vishwanath}}, \ and\
  \bibinfo {author} {\bibfnamefont {X.}~\bibnamefont {Wan}},\ }\href@noop {}
  {\bibfield  {journal} {\bibinfo  {journal} {Nature}\ }\textbf {\bibinfo
  {volume} {566}},\ \bibinfo {pages} {486} (\bibinfo {year}
  {2019})}\BibitemShut {NoStop}%
\bibitem [{\citenamefont {Song}\ \emph {et~al.}(2019)\citenamefont {Song},
  \citenamefont {Elcoro}, \citenamefont {Regnault},\ and\ \citenamefont
  {Bernevig}}]{son19}%
  \BibitemOpen
  \bibfield  {author} {\bibinfo {author} {\bibfnamefont {Z.}~\bibnamefont
  {Song}}, \bibinfo {author} {\bibfnamefont {L.}~\bibnamefont {Elcoro}},
  \bibinfo {author} {\bibfnamefont {N.}~\bibnamefont {Regnault}}, \ and\
  \bibinfo {author} {\bibfnamefont {B.~A.}\ \bibnamefont {Bernevig}},\
  }\href@noop {} {\bibfield  {journal} {\bibinfo  {journal} {arXiv e-prints}\
  ,\ \bibinfo {pages} {arXiv:1905.03262}} (\bibinfo {year} {2019})}\BibitemShut
  {NoStop}%
\bibitem [{\citenamefont {Po}\ \emph {et~al.}(2018)\citenamefont {Po},
  \citenamefont {Watanabe},\ and\ \citenamefont {Vishwanath}}]{po18}%
  \BibitemOpen
  \bibfield  {author} {\bibinfo {author} {\bibfnamefont {H.~C.}\ \bibnamefont
  {Po}}, \bibinfo {author} {\bibfnamefont {H.}~\bibnamefont {Watanabe}}, \ and\
  \bibinfo {author} {\bibfnamefont {A.}~\bibnamefont {Vishwanath}},\
  }\href@noop {} {\bibfield  {journal} {\bibinfo  {journal} {Phys. Rev. Lett.}\
  }\textbf {\bibinfo {volume} {121}},\ \bibinfo {pages} {126402} (\bibinfo
  {year} {2018})}\BibitemShut {NoStop}%
\bibitem [{\citenamefont {Bradlyn}\ \emph {et~al.}(2019)\citenamefont
  {Bradlyn}, \citenamefont {Wang}, \citenamefont {Cano},\ and\ \citenamefont
  {Bernevig}}]{bra19}%
  \BibitemOpen
  \bibfield  {author} {\bibinfo {author} {\bibfnamefont {B.}~\bibnamefont
  {Bradlyn}}, \bibinfo {author} {\bibfnamefont {Z.}~\bibnamefont {Wang}},
  \bibinfo {author} {\bibfnamefont {J.}~\bibnamefont {Cano}}, \ and\ \bibinfo
  {author} {\bibfnamefont {B.~A.}\ \bibnamefont {Bernevig}},\ }\href@noop {}
  {\bibfield  {journal} {\bibinfo  {journal} {Phys. Rev. B}\ }\textbf {\bibinfo
  {volume} {99}},\ \bibinfo {pages} {045140} (\bibinfo {year}
  {2019})}\BibitemShut {NoStop}%
\bibitem [{\citenamefont {Cao}\ \emph {et~al.}(2018{\natexlab{a}})\citenamefont
  {Cao}, \citenamefont {Fatemi}, \citenamefont {Fang}, \citenamefont
  {Watanabe}, \citenamefont {Taniguchi}, \citenamefont {Kaxiras},\ and\
  \citenamefont {Jarillo-Herrero}}]{cao2018}%
  \BibitemOpen
  \bibfield  {author} {\bibinfo {author} {\bibfnamefont {Y.}~\bibnamefont
  {Cao}}, \bibinfo {author} {\bibfnamefont {V.}~\bibnamefont {Fatemi}},
  \bibinfo {author} {\bibfnamefont {S.}~\bibnamefont {Fang}}, \bibinfo {author}
  {\bibfnamefont {K.}~\bibnamefont {Watanabe}}, \bibinfo {author}
  {\bibfnamefont {T.}~\bibnamefont {Taniguchi}}, \bibinfo {author}
  {\bibfnamefont {E.}~\bibnamefont {Kaxiras}}, \ and\ \bibinfo {author}
  {\bibfnamefont {P.}~\bibnamefont {Jarillo-Herrero}},\ }\href@noop {}
  {\bibfield  {journal} {\bibinfo  {journal} {Nature}\ }\textbf {\bibinfo
  {volume} {556}},\ \bibinfo {pages} {43} (\bibinfo {year}
  {2018}{\natexlab{a}})}\BibitemShut {NoStop}%
\bibitem [{\citenamefont {Cao}\ \emph {et~al.}(2018{\natexlab{b}})\citenamefont
  {Cao}, \citenamefont {Fatemi}, \citenamefont {Demir}, \citenamefont {Fang},
  \citenamefont {Tomarken}, \citenamefont {Luo}, \citenamefont
  {Sanchez-Yamagishi}, \citenamefont {Watanabe}, \citenamefont {Taniguchi},
  \citenamefont {Kaxiras} \emph {et~al.}}]{cao18b}%
  \BibitemOpen
  \bibfield  {author} {\bibinfo {author} {\bibfnamefont {Y.}~\bibnamefont
  {Cao}}, \bibinfo {author} {\bibfnamefont {V.}~\bibnamefont {Fatemi}},
  \bibinfo {author} {\bibfnamefont {A.}~\bibnamefont {Demir}}, \bibinfo
  {author} {\bibfnamefont {S.}~\bibnamefont {Fang}}, \bibinfo {author}
  {\bibfnamefont {S.~L.}\ \bibnamefont {Tomarken}}, \bibinfo {author}
  {\bibfnamefont {J.~Y.}\ \bibnamefont {Luo}}, \bibinfo {author} {\bibfnamefont
  {J.~D.}\ \bibnamefont {Sanchez-Yamagishi}}, \bibinfo {author} {\bibfnamefont
  {K.}~\bibnamefont {Watanabe}}, \bibinfo {author} {\bibfnamefont
  {T.}~\bibnamefont {Taniguchi}}, \bibinfo {author} {\bibfnamefont
  {E.}~\bibnamefont {Kaxiras}},  \emph {et~al.},\ }\href@noop {} {\bibfield
  {journal} {\bibinfo  {journal} {Nature}\ }\textbf {\bibinfo {volume} {556}},\
  \bibinfo {pages} {80} (\bibinfo {year} {2018}{\natexlab{b}})}\BibitemShut
  {NoStop}%
\bibitem [{\citenamefont {Ahn}\ \emph {et~al.}(2019)\citenamefont {Ahn},
  \citenamefont {Park},\ and\ \citenamefont {Yang}}]{ahn19}%
  \BibitemOpen
  \bibfield  {author} {\bibinfo {author} {\bibfnamefont {J.}~\bibnamefont
  {Ahn}}, \bibinfo {author} {\bibfnamefont {S.}~\bibnamefont {Park}}, \ and\
  \bibinfo {author} {\bibfnamefont {B.-J.}\ \bibnamefont {Yang}},\ }\href
  {\doibase 10.1103/PhysRevX.9.021013} {\bibfield  {journal} {\bibinfo
  {journal} {Phys. Rev. X}\ }\textbf {\bibinfo {volume} {9}},\ \bibinfo {pages}
  {021013} (\bibinfo {year} {2019})}\BibitemShut {NoStop}%
\bibitem [{\citenamefont {Po}\ \emph {et~al.}(2019)\citenamefont {Po},
  \citenamefont {Zou}, \citenamefont {Senthil},\ and\ \citenamefont
  {Vishwanath}}]{po19}%
  \BibitemOpen
  \bibfield  {author} {\bibinfo {author} {\bibfnamefont {H.~C.}\ \bibnamefont
  {Po}}, \bibinfo {author} {\bibfnamefont {L.}~\bibnamefont {Zou}}, \bibinfo
  {author} {\bibfnamefont {T.}~\bibnamefont {Senthil}}, \ and\ \bibinfo
  {author} {\bibfnamefont {A.}~\bibnamefont {Vishwanath}},\ }\href@noop {}
  {\bibfield  {journal} {\bibinfo  {journal} {Phys. Rev. B}\ }\textbf {\bibinfo
  {volume} {99}},\ \bibinfo {pages} {195455} (\bibinfo {year}
  {2019})}\BibitemShut {NoStop}%
\bibitem [{\citenamefont {Zak}(1982)}]{zak82}%
  \BibitemOpen
  \bibfield  {author} {\bibinfo {author} {\bibfnamefont {J.}~\bibnamefont
  {Zak}},\ }\href {\doibase 10.1103/PhysRevB.26.3010} {\bibfield  {journal}
  {\bibinfo  {journal} {Phys. Rev. B}\ }\textbf {\bibinfo {volume} {26}},\
  \bibinfo {pages} {3010} (\bibinfo {year} {1982})}\BibitemShut {NoStop}%
\bibitem [{\citenamefont {Lau}\ \emph {et~al.}(2016)\citenamefont {Lau},
  \citenamefont {van~den Brink},\ and\ \citenamefont {Ortix}}]{lau16}%
  \BibitemOpen
  \bibfield  {author} {\bibinfo {author} {\bibfnamefont {A.}~\bibnamefont
  {Lau}}, \bibinfo {author} {\bibfnamefont {J.}~\bibnamefont {van~den Brink}},
  \ and\ \bibinfo {author} {\bibfnamefont {C.}~\bibnamefont {Ortix}},\
  }\href@noop {} {\bibfield  {journal} {\bibinfo  {journal} {Phys. Rev. B}\
  }\textbf {\bibinfo {volume} {94}},\ \bibinfo {pages} {165164} (\bibinfo
  {year} {2016})}\BibitemShut {NoStop}%
\bibitem [{\citenamefont {van Miert}\ and\ \citenamefont
  {Ortix}(2017)}]{mie17}%
  \BibitemOpen
  \bibfield  {author} {\bibinfo {author} {\bibfnamefont {G.}~\bibnamefont {van
  Miert}}\ and\ \bibinfo {author} {\bibfnamefont {C.}~\bibnamefont {Ortix}},\
  }\href@noop {} {\bibfield  {journal} {\bibinfo  {journal} {Phys. Rev. B}\
  }\textbf {\bibinfo {volume} {96}},\ \bibinfo {pages} {235130} (\bibinfo
  {year} {2017})}\BibitemShut {NoStop}%
\bibitem [{\citenamefont {Fu}\ and\ \citenamefont {Kane}(2006)}]{fu06}%
  \BibitemOpen
  \bibfield  {author} {\bibinfo {author} {\bibfnamefont {L.}~\bibnamefont
  {Fu}}\ and\ \bibinfo {author} {\bibfnamefont {C.~L.}\ \bibnamefont {Kane}},\
  }\href {\doibase 10.1103/PhysRevB.74.195312} {\bibfield  {journal} {\bibinfo
  {journal} {Phys. Rev. B}\ }\textbf {\bibinfo {volume} {74}},\ \bibinfo
  {pages} {195312} (\bibinfo {year} {2006})}\BibitemShut {NoStop}%
\bibitem [{\citenamefont {Alexandradinata}\ \emph {et~al.}(2014)\citenamefont
  {Alexandradinata}, \citenamefont {Dai},\ and\ \citenamefont
  {Bernevig}}]{ale14}%
  \BibitemOpen
  \bibfield  {author} {\bibinfo {author} {\bibfnamefont {A.}~\bibnamefont
  {Alexandradinata}}, \bibinfo {author} {\bibfnamefont {X.}~\bibnamefont
  {Dai}}, \ and\ \bibinfo {author} {\bibfnamefont {B.~A.}\ \bibnamefont
  {Bernevig}},\ }\href@noop {} {\bibfield  {journal} {\bibinfo  {journal}
  {Phys. Rev. B}\ }\textbf {\bibinfo {volume} {89}},\ \bibinfo {pages} {155114}
  (\bibinfo {year} {2014})}\BibitemShut {NoStop}%
\bibitem [{\citenamefont {{Kruthoff}}\ \emph {et~al.}(2017)\citenamefont
  {{Kruthoff}}, \citenamefont {{de Boer}},\ and\ \citenamefont {{van
  Wezel}}}]{kru17}%
  \BibitemOpen
  \bibfield  {author} {\bibinfo {author} {\bibfnamefont {J.}~\bibnamefont
  {{Kruthoff}}}, \bibinfo {author} {\bibfnamefont {J.}~\bibnamefont {{de
  Boer}}}, \ and\ \bibinfo {author} {\bibfnamefont {J.}~\bibnamefont {{van
  Wezel}}},\ }\href@noop {} {\bibfield  {journal} {\bibinfo  {journal} {arXiv
  e-prints}\ ,\ \bibinfo {pages} {arXiv:1711.04769}} (\bibinfo {year}
  {2017})}\BibitemShut {NoStop}%
\bibitem [{\citenamefont {van Miert}\ and\ \citenamefont
  {Ortix}(2018{\natexlab{b}})}]{mie18r}%
  \BibitemOpen
  \bibfield  {author} {\bibinfo {author} {\bibfnamefont {G.}~\bibnamefont {van
  Miert}}\ and\ \bibinfo {author} {\bibfnamefont {C.}~\bibnamefont {Ortix}},\
  }\href {\doibase 10.1103/PhysRevB.97.201111} {\bibfield  {journal} {\bibinfo
  {journal} {Phys. Rev. B}\ }\textbf {\bibinfo {volume} {97}},\ \bibinfo
  {pages} {201111} (\bibinfo {year} {2018}{\natexlab{b}})}\BibitemShut
  {NoStop}%
\bibitem [{\citenamefont {Cano}\ \emph {et~al.}(2018)\citenamefont {Cano},
  \citenamefont {Bradlyn}, \citenamefont {Wang}, \citenamefont {Elcoro},
  \citenamefont {Vergniory}, \citenamefont {Felser}, \citenamefont {Aroyo},\
  and\ \citenamefont {Bernevig}}]{can18}%
  \BibitemOpen
  \bibfield  {author} {\bibinfo {author} {\bibfnamefont {J.}~\bibnamefont
  {Cano}}, \bibinfo {author} {\bibfnamefont {B.}~\bibnamefont {Bradlyn}},
  \bibinfo {author} {\bibfnamefont {Z.}~\bibnamefont {Wang}}, \bibinfo {author}
  {\bibfnamefont {L.}~\bibnamefont {Elcoro}}, \bibinfo {author} {\bibfnamefont
  {M.~G.}\ \bibnamefont {Vergniory}}, \bibinfo {author} {\bibfnamefont
  {C.}~\bibnamefont {Felser}}, \bibinfo {author} {\bibfnamefont {M.~I.}\
  \bibnamefont {Aroyo}}, \ and\ \bibinfo {author} {\bibfnamefont {B.~A.}\
  \bibnamefont {Bernevig}},\ }\href@noop {} {\bibfield  {journal} {\bibinfo
  {journal} {Phys. Rev. Lett.}\ }\textbf {\bibinfo {volume} {120}},\ \bibinfo
  {pages} {266401} (\bibinfo {year} {2018})}\BibitemShut {NoStop}%
\bibitem [{Note1()}]{Note1}%
  \BibitemOpen
  \bibinfo {note} {A real gauge can be formulated as $\protect \mathcal
  {C}_{2}\Theta \ket {\psi }=\ket {\psi }$.}\BibitemShut {Stop}%
\bibitem [{\citenamefont {Bzdu\ifmmode~\check{s}\else \v{s}\fi{}ek}\ and\
  \citenamefont {Sigrist}(2017)}]{bzd17}%
  \BibitemOpen
  \bibfield  {author} {\bibinfo {author} {\bibfnamefont {T.}~\bibnamefont
  {Bzdu\ifmmode~\check{s}\else \v{s}\fi{}ek}}\ and\ \bibinfo {author}
  {\bibfnamefont {M.}~\bibnamefont {Sigrist}},\ }\href@noop {} {\bibfield
  {journal} {\bibinfo  {journal} {Phys. Rev. B}\ }\textbf {\bibinfo {volume}
  {96}},\ \bibinfo {pages} {155105} (\bibinfo {year} {2017})}\BibitemShut
  {NoStop}%
\bibitem [{\citenamefont {Ahn}\ \emph {et~al.}(2018)\citenamefont {Ahn},
  \citenamefont {Kim}, \citenamefont {Kim},\ and\ \citenamefont
  {Yang}}]{ahn18}%
  \BibitemOpen
  \bibfield  {author} {\bibinfo {author} {\bibfnamefont {J.}~\bibnamefont
  {Ahn}}, \bibinfo {author} {\bibfnamefont {D.}~\bibnamefont {Kim}}, \bibinfo
  {author} {\bibfnamefont {Y.}~\bibnamefont {Kim}}, \ and\ \bibinfo {author}
  {\bibfnamefont {B.-J.}\ \bibnamefont {Yang}},\ }\href@noop {} {\bibfield
  {journal} {\bibinfo  {journal} {Phys. Rev. Lett.}\ }\textbf {\bibinfo
  {volume} {121}},\ \bibinfo {pages} {106403} (\bibinfo {year}
  {2018})}\BibitemShut {NoStop}%
\bibitem [{\citenamefont {Franca}\ \emph {et~al.}(2018)\citenamefont {Franca},
  \citenamefont {van~den Brink},\ and\ \citenamefont {Fulga}}]{fra18}%
  \BibitemOpen
  \bibfield  {author} {\bibinfo {author} {\bibfnamefont {S.}~\bibnamefont
  {Franca}}, \bibinfo {author} {\bibfnamefont {J.}~\bibnamefont {van~den
  Brink}}, \ and\ \bibinfo {author} {\bibfnamefont {I.~C.}\ \bibnamefont
  {Fulga}},\ }\href@noop {} {\bibfield  {journal} {\bibinfo  {journal} {Phys.
  Rev. B}\ }\textbf {\bibinfo {volume} {98}},\ \bibinfo {pages} {201114}
  (\bibinfo {year} {2018})}\BibitemShut {NoStop}%
\bibitem [{Note2()}]{Note2}%
  \BibitemOpen
  \bibinfo {note} {Since we have to take a region symmetrically centered around
  $\nu =0,1/2$ we have to either include both or neither of the blue
  bands.}\BibitemShut {Stop}%
\bibitem [{\citenamefont {Harper}(1955)}]{har55}%
  \BibitemOpen
  \bibfield  {author} {\bibinfo {author} {\bibfnamefont {P.~G.}\ \bibnamefont
  {Harper}},\ }\href {http://stacks.iop.org/0370-1298/68/i=10/a=304} {\bibfield
   {journal} {\bibinfo  {journal} {Proceedings of the Physical Society. Section
  A}\ }\textbf {\bibinfo {volume} {68}},\ \bibinfo {pages} {874} (\bibinfo
  {year} {1955})}\BibitemShut {NoStop}%
\bibitem [{\citenamefont {Aubry}\ and\ \citenamefont {Andr\'e}(1980)}]{aub80}%
  \BibitemOpen
  \bibfield  {author} {\bibinfo {author} {\bibfnamefont {S.}~\bibnamefont
  {Aubry}}\ and\ \bibinfo {author} {\bibfnamefont {G.}~\bibnamefont
  {Andr\'e}},\ }\href@noop {} {\bibfield  {journal} {\bibinfo  {journal} {Ann.
  Isr. Phys. Soc.}\ }\textbf {\bibinfo {volume} {3}},\ \bibinfo {pages} {133}
  (\bibinfo {year} {1980})}\BibitemShut {NoStop}%
\bibitem [{\citenamefont {Ganeshan}\ \emph {et~al.}(2013)\citenamefont
  {Ganeshan}, \citenamefont {Sun},\ and\ \citenamefont {Das~Sarma}}]{gan13}%
  \BibitemOpen
  \bibfield  {author} {\bibinfo {author} {\bibfnamefont {S.}~\bibnamefont
  {Ganeshan}}, \bibinfo {author} {\bibfnamefont {K.}~\bibnamefont {Sun}}, \
  and\ \bibinfo {author} {\bibfnamefont {S.}~\bibnamefont {Das~Sarma}},\ }\href
  {\doibase 10.1103/PhysRevLett.110.180403} {\bibfield  {journal} {\bibinfo
  {journal} {Phys. Rev. Lett.}\ }\textbf {\bibinfo {volume} {110}},\ \bibinfo
  {pages} {180403} (\bibinfo {year} {2013})}\BibitemShut {NoStop}%
\end{thebibliography}
\end{document}